\documentclass[12pt]{article}
\usepackage{hyperref}
\usepackage{cite}
\usepackage{color}
\usepackage{graphicx}
\usepackage{amsmath}
\usepackage{amssymb}
\usepackage{xspace}

\makeatletter
\@addtoreset{equation}{section}

\makeatletter
\renewcommand\section{\@startsection {section}{1}{\z@}%
                                   {-3.5ex \@plus -1ex \@minus -.2ex}
                                   {2.3ex \@plus.2ex}%
                                   {\normalfont\large\bfseries}}
\renewcommand\subsection{\@startsection{subsection}{2}{\z@}%
                                     {-3.25ex\@plus -1ex \@minus -.2ex}%
                                     {1.5ex \@plus .2ex}%
                                     {\normalfont\bfseries}}

\def\baselinestretch{1.2}
\newcommand{\be}{\begin{equation}}
\newcommand{\ee}{\end{equation}}
\newcommand{\beq}{\begin{eqnarray}}
\newcommand{\eeq}{\end{eqnarray}}

\newcommand{\tr}{{\rm Tr}}
\newcommand{\gone}[1]{{}}
\newcommand{\bl}{\noindent $\bullet\ $}

\begin{document}
\begin{titlepage}
\begin{flushright}
MAD-TH-09-05, WIS/04/09-JUN-DPP
\end{flushright}

\vfil

\begin{center}

{\bf \Large
D-brane Charges in Gravitational Duals of $2+1$}

{\bf \Large Dimensional
Gauge Theories and Duality Cascades
}

\vfil

Ofer Aharony$^a$, Akikazu Hashimoto$^b$, Shinji Hirano$^c$, and Peter Ouyang$^b$

\vfil

$^a$ Department of Particle Physics, The Weizmann Institute of Science, Rehovot 76100,
Israel

$^b$ Department of Physics, University of Wisconsin, Madison, WI 53706, USA

$^c$ The Niels Bohr Institute, Blegdamsvej 17, DK-2100 Copenhagen, Denmark

\vfil

\end{center}

\begin{abstract}
\noindent
We perform a systematic analysis of the D-brane charges associated
with string theory realizations of $d=3$ gauge theories, focusing on
the examples of the ${\cal N}=4$ supersymmetric $U(N)\times U(N+M)$
Yang-Mills theory and the ${\cal N}=3$ supersymmetric $U(N)\times
U(N+M)$ Yang-Mills-Chern-Simons theory. We use both the brane
construction of these theories and their dual string theory
backgrounds in the supergravity approximation. In the ${\cal N}=4$
case we generalize the previously known gravitational duals to
arbitrary values of the gauge couplings, and present a precise mapping
between the gravity and field theory parameters. In the ${\cal N}=3$
case, which (for some values of $N$ and $M$) flows to an ${\cal N}=6$
supersymmetric Chern-Simons-matter theory in the IR, we argue that the
careful analysis of the charges leads to a shift in the value of the
$B_2$ field in the IR solution by $1/2$, in units where its
periodicity is one, compared to previous claims. We also suggest that
the ${\cal N}=3$ theories may exhibit, for some values of $N$ and $M$,
duality cascades similar to those of the Klebanov-Strassler theory.

\end{abstract}
\vspace{0.5in}

\end{titlepage}
\renewcommand{\baselinestretch}{1.05}  

\section{Introduction}

There has been significant recent progress in the study of
superconformal field theories in $2+1$ dimensions, following the
discovery of a Lagrangian for a theory with ${\cal N}=8$ supersymmetry
by Bagger, Lambert, and Gustavsson
\cite{Bagger:2007jr,Gustavsson:2007vu}.  The discovery of this theory
is significant in light of the fact that a Lagrangian formulation for
theories with these properties was believed not to exist
\cite{Schwarz:2004yj}. The original formulation of these new models
used an exotic mathematical structure called a ``3-algebra.''
Subsequently, it was shown that these theories admit a formulation in
terms of an ordinary Chern-Simons gauge theory with an $SU(2)\times
SU(2)$ gauge group and bi-fundamental matter
\cite{VanRaamsdonk:2008ft}.  These theories are closely related to the
decoupled theories living on the worldvolume of M2-branes in M theory.

An interesting class of generalizations with ${\cal N}=6$
supersymmetry, that describes the decoupled theory on $N_2$ M2-branes
at a ${\bf C}^4/{\bf Z}_k$ orbifold point, and that has a known
gravitational dual, was identified by Aharony, Bergman, Jafferis, and
Maldacena (ABJM) \cite{Aharony:2008ug}.  It arises from a brane
construction consisting of $N_2$ D3-branes on ${\bf R}^{1,2} \times
S^1$ intersecting with NS5 and $(1,k)$ 5-brane defects.  After taking
a decoupling limit, the result is a Chern-Simons theory with gauge
group $U(N) \times U(N)$ with $N = N_2$ and with Chern-Simons terms of
levels $k$ and $-k$ for the two gauge groups, coupled to scalar and
fermion bifundamental fields.  The supersymmetry is enhanced to ${\cal
N}=8$ for $k=1,2$. The theory has a dual gravitational formulation as
M theory on a ${\bf Z}_k$ orbifold of $AdS_4 \times S^7$.  For
sufficiently large values of $k$, it is natural to describe this model
in a type IIA description by reducing along the Hopf fiber of the
$S^7$. The background geometry is then $AdS_4 \times {\bf CP}^3$, with
some RR fluxes.

A generalization of ABJM adding fractional branes, with gauge group
$U(N) \times U(N+M)$ (where $N=N_2$, $M=N_4$ and $N_4 \leq k$), was
considered by Aharony, Bergman, and Jafferis (ABJ)
\cite{Aharony:2008gk}. According to ABJ, the dual M theory description
of this model is the same $AdS_4 \times S^7/{\bf Z}_k$ geometry, with
$N_4/k$ units of discrete (torsion) 3-form flux through an $S^3/{\bf
Z}_k$ cycle in $S^7/{\bf Z}_k$. The type IIA description of this
background was claimed to be $AdS_4 \times \bf{CP}^3$, with an NS-NS
2-form field through the $\bf{CP}^1$ 2-cycle in ${\bf CP}^3$, equal to
$b = -N_4 / k$ (in units where its periodicity is one\footnote{The
minus sign is the result of the conventions used in this paper,
summarized in appendix \ref{appB}.}).

Several consistency tests for this conjecture were presented in
\cite{Aharony:2008gk}.  There are, however, some subtle and confusing
aspects to the proposal of ABJ. In most constructions of gravity duals
involving product gauge groups, the flux of the $B_2$ field through
the 2-cycle is continuous.  In the brane construction of these
theories \cite{Klebanov:1998hh,Morrison:1998cs} it parameterizes the
distance between NS5-branes, which is directly related to the relative
strength of the gauge couplings. The fact that the gauge couplings
have decoupled by the time one flows to the superconformal theory in
the IR assures us that there is no immediate contradiction with the
discreteness of $b$ in the supergravity dual of the IR fixed
point. Nonetheless, it would be interesting to understand how this
decoupling manifests itself in the dual gravitational description of
the renormalization group (RG) flow, when the superconformal field
theory in the IR is embedded into a supersymmetric
Yang-Mills-Chern-Simons-Matter theory. It appears that some form of
attractor mechanism is at work, with $b$ flowing from a continuous
value in the UV to a discrete value in the IR.

Another slightly counterintuitive aspect of the proposal of ABJ is
that the flux of the {\it potential field} $B_2$, rather than that of
its {\it field strength}, is quantized and related to integer charges
characterizing the system. This is not an entirely unfamiliar notion.
Similar quantizations of potentials also appear in the near-horizon
limit of a D1-D3 bound state for a D3-brane wrapping a $T^2$
\cite{Hashimoto:1999yj}. The general mechanism for quantization of a
gauge potential arises when there are Chern-Simons terms in the
action, mixing gauge potentials and gauge field strengths. Such
Chern-Simons terms are certainly part of supergravities in 10 and 11
dimensions.

Chern-Simons terms introduce various subtleties in measuring and
quantizing charges.  In simple situations, we usually expect charges
to satisfy all of the following four properties:
\begin{itemize}
\item Localization: we expect elementary sources of charge to be
point-like (or localized on branes) and to respect Gauss' law.
\item Gauge invariance: we expect charge to be a physical,
gauge-invariant notion.
\item Quantization: we expect charges to respect a Dirac quantization
condition.
\item
Conservation: we expect charges to be an invariant quantity in
dynamical processes.
\end{itemize}
As reviewed by Marolf \cite{Marolf:2000cb}, in the presence of
Chern-Simons terms there is in general no definition of charge which
simultaneously respects all four of the properties listed above.
Instead, there are three different notions of charge, which Marolf
called Maxwell, Page, and Brane charges, each of which satisfies a
subset of the properties above, and which can take different values.
So, care is needed when discussing issues such as quantization of
charges and enumeration of gauge equivalence classes of physical
configurations.

In this article, we will closely examine the D-brane charges and the
related quantization of the potential field in the models of ABJM and
ABJ \cite{Aharony:2008ug,Aharony:2008gk}, and also in the ${\cal N}=3$
gauge theories obtained by adding Yang-Mills kinetic terms to the ABJM
and ABJ theories. Most of the relevant issues arise already when
looking at the same theories with $k=0$ (no Chern-Simons coupling),
which leads to ${\cal N}=4$ supersymmetric gauge theories.  Section 2
is devoted to analyzing this case (including also the possibility of
having additional hypermultiplets in the fundamental representation of
one of the gauge groups).  In section 3, we investigate the
quantization conditions of charges and potentials in the ABJM and ABJ
models.  Broadly speaking, we confirm the basic picture of ABJ and
ABJM where the gauge potentials take quantized values in the IR,
depending on the quantized charges. However, we find a subtle
half-integer shift in $b$, related to the Freed-Witten anomaly
\cite{Freed:1999vc}, which corrects the specific quantized value of
the gauge potential field.\footnote{This shift also contributes to the
correction to the radius of $AdS_4$ due to the effects of curvature
corrections and discrete torsion discussed in \cite{Bergman:2009zh}.}
We verify that the consistency tests carried out in ABJ and ABJM are
compatible with this correction.

Most of the subtleties in charge quantization are already present in
the D1-D3 bound state system in type IIB string theory with some NS-NS
$B_2$-field in the background. We review this setup in appendix
\ref{appA}.

The same subtleties in defining charges appear also in the
Klebanov-Strassler theory \cite{Klebanov:2000hb}, where they are an
essential part of a ``duality cascade.'' In the corresponding
gravitational backgrounds, the D3-brane Maxwell charge changes during
the RG flow (corresponding to a reduction in the rank of the gauge
group when flowing to the IR), while the Page charge is quantized. The
Page charge is not invariant under large gauge transformations, and
these transformations may be viewed as Seiberg duality transformations
that relate the IR behavior of different field theories. The cascade
of \cite{Klebanov:2000hb} may be viewed as a sequence of such duality
transformations. The ${\cal N}=3$ theories discussed above also have a
Seiberg-like duality in the IR \cite{Aharony:2008gk}, which in the
brane construction looks very similar to the one of
\cite{Klebanov:2000hb}; thus, it is natural to suggest that these
theories may also exhibit duality cascades. We discuss this
possibility, and its implications for the IR behavior of theories with
general values of $N_2$ and $N_4$, in section 4. Appendix \ref{appB}
contains a summary of our conventions.

\section{A $d=3$ ${\cal N}=4$ theory with fractional branes and flavors}

\subsection{The dual gravitational background}

An interesting $d=3$ gauge theory with a known gravitational dual is
the ${\cal N}=4$ supersymmetric gauge theory arising from the
low-energy theory on $N_2$ D2-branes and $N_6$ D6-branes in type IIA
string theory, oriented as
\be \begin{tabular}{lcccccccccc}
& 0 & 1 & 2 & 3& 4 & 5 & 6& 7&8&9 \\
D2 & $\bullet$ & $\bullet$ & $\bullet$ \\
D6 & $\bullet$ & $\bullet$ & $\bullet$ & $\bullet$ & $\bullet$ & $\bullet$ & $\bullet$ &
\end{tabular} \label{orientation}
\ee
where the coordinates 3456 span an ALE space on which the D6-branes
are wrapped. This configuration preserves eight supercharges, and when
lifted to M theory it describes $N_2$ M2-branes localized in an
8-dimensional transverse geometry of the form $ALE \times TN_k$, with
$k = N_6$ \cite{Cherkis:2002ir}. The action for supergravity in 11
dimensions takes the standard form\footnote{See Appendix \ref{appB}
for a summary of our conventions.}
\be S_{11} = {1 \over 2 \kappa_{11}^2} \int d^{11}x\,  \sqrt{-g} \left(R - {1 \over 2}
|G_4|^2 \right) - {1 \over  2 \kappa_{11}^2} \int {1 \over 6}   C_3 \wedge G_4 \wedge G_4,
\ee
which can be solved by an ansatz of the form
\beq ds^2 &=& H^{-2/3} (-dt^2+dx_1^2+dx_2^2) + H^{1/3} (ds_{ALE}^2 +ds_{TN_k}^2), \\
G_4 &=& dC_3 = dt \wedge dx_1 \wedge dx_2 \wedge d H^{-1},
\label{ansatz1}\eeq
where the Taub-NUT metric is given by
\be ds_{TN_k}^2 = V(r)^{-1} (d r^2 + r^2 (d \theta^2 + \sin^2\theta d \phi^2)) +V(r)
R_{11}^2 k^2
\left( d \psi - {1 \over 2}\cos \theta d \phi \right)^2 \label{TNmetric} \ee
with
\be V(r) \equiv
\left(1 + {k R_{11} \over 2 r} \right)^{-1}, \qquad R_{11} = g_s l_s\ , \ee
for the range of coordinates $0 \le r < \infty$, $0 \le \theta \le
\pi$, $0 \le \phi \le 2 \pi$, $0 \le \psi \le 2 \pi/k$.  The field
equations imply that the warp factor $H$ is a solution of the Poisson
equation
\be \nabla^2_{TN} H(\vec y,\vec r) + \nabla^2_y H(\vec y,\vec r) = -(2 \pi  l_p)^6 N_2
\delta^4(\vec y) \delta^4(\vec r) \ , \ee
where $\vec{y}$ denotes the ALE coordinates and $\vec{r}$ denotes the
Taub-NUT coordinates, with a source coming from $N_2$ M2-branes at the
origin\footnote{In this section we assume that the charges are large,
and we neglect the corrections to the charges coming from the
curvature.}.  This equation is separable in $y$ and $r$, where $y$ is
the radial variable of the ALE space, and the solution can be written
in the form of a convolution of a Bessel function and a confluent
hypergeometric function \cite{Cherkis:2002ir}.  If we take the ALE
space to be the ${\bf R}^4 / {\bf Z}_2$ orbifold, as we will assume
from here on, this geometry describes the supergravity dual of a
$U(N_2) \times U(N_2)$ gauge theory, with two bifundamental
hypermultiplets and with $k=N_6$ hypermultiplets in the fundamental
representation of one of the $U(N_2)$ gauge groups. (The issue of
which one of the two will be discussed below.) This is easiest to see
in the T-dual brane construction, that we will describe in section
\ref{brane_nfour} below. Of course, the M theory description is only
useful at low energies, but the same solutions (reduced on the M
theory circle, as we will describe below) are valid also in type IIA
supergravity for higher energies. This class of backgrounds is
particularly useful to visualize the Intriligator-Seiberg mirror
symmetry \cite{Intriligator:1996ex} for field theories in $2+1$
dimensions from the point of view of the holographic dual\footnote{One
can also consider taking a $TN_{k1} \times TN_{k2}$
space-time. Reducing to type IIA and T-dualizing implies that this
corresponds to a field theory on $N_2$ D3-branes intersecting $k_1$
NS5-branes and $k_2$ D5-branes, although the supergravity description
only captures the aspect of the theory where the positions of the
defects are smeared. There is a decoupling limit in which this
describes a $3+1$ dimensional gauge theory with defects at the
positions of the brane intersections, and a further decoupling limit
where we reduce to the low-energy $2+1$ dimensional gauge
theory.\label{fn}}\cite{Cherkis:2002ir}.

Generalizing this construction to the case where we include also $N_4$
fractional branes, which are (in the type IIA description) D4-branes
wrapped on the vanishing 2-cycle of the ALE space, is relatively
straightforward. If we are interested in the case where all of the D2,
D4, and D6-branes are coincident in the type IIA description, we can
take the ansatz
\beq ds^2 &=& H^{-2/3} (-dt^2+dx_1^2+dx_2^2) + H^{1/3} (ds_{ALE}^2+ds_{TN_k}^2), \\
G_4 &=& dC_3 =  dt \wedge dx_1 \wedge dx_2 \wedge d H^{-1} + G_4^{SD}, \label{g4}\\
G_4^{SD} &=& d(l V \omega_2 \wedge \sigma_3 + 2 \alpha \omega_2 \wedge d \psi) \eeq
for some constants $l$ and $\alpha$, where
\be {1 \over 2} \sigma_3 \equiv d \psi - {1 \over 2} \cos \theta d \phi \ , \ee
and $\omega_2$ is the self-dual 2-form dual to the collapsed 2-cycle
in the ALE space, normalized so that\footnote{This convention is used in
\cite{Grana:2001xn}.}
\be \int_{ALE} \omega_2 \wedge \omega_2 = {1 \over 2} \ . \ee
We denote the 2-cycle dual to $\omega_2$ by $\Sigma$, and the
non-compact 2-cycle which is dual to $\Sigma$ on the ALE space by
${\cal M}$. Then, it follows that
\be \int_\Sigma \omega_2 = 1, \qquad \int_{{\cal M}} \omega_2 = {1
\over 2} \ . \ee
The four-form $G_4^{SD}$ is a self-dual 4-form on $ALE\times TN_k$. We
will fix the parameters $l$ and $\alpha$ in terms of the field theory
data shortly.

To express this ansatz in terms of type IIA supergravity, let $d
x_{11} = k R_{11} d\psi$ and perform the standard reduction. The
ansatz can then be expressed in the form
\beq ds_{IIA}^2 &=& H^{-1/2} V^{1/2} (-dt^2 + dx_1^2+dx_2^2) + H^{1/2} V^{1/2} ds_{ALE}^2
\label{iiaf} \cr
&& \qquad  + H^{1/2} V^{-1/2} (dr^2 + r^2 (d \theta^2 + \sin^2 \theta d \phi^2)),\\
A_1 & = & -{1 \over 2} R_{11}k  \cos \theta d \phi, \\
A_3 & = & -(H^{-1}-1) dt \wedge dx_1 \wedge dx_2 - l V \omega_2 \wedge \cos \theta d
\phi,\\
B_2 & = &   -{2 \over R_{11} k} (l V \omega_2 +  \alpha \omega_2),  \\
e^{\phi} &=& g_s  H^{1/4} V^{3/4}. \label{iial}
\eeq
It is convenient to introduce a field variable $b$ by the relation
\be B_2 = (2 \pi)^2 \alpha' b\,  \omega_2, \ee
such that a large gauge transformation of $B_2$ takes $b \to b+1$.

The equations of motion of 11 dimensional supergravity are satisfied
provided $H$ obeys
\be 0 =  \left(\nabla^2_y+ \nabla^2_{TN} \right) H + {l^2 V^4\over 2  r^4}  \delta^4(\vec y)
+ (2 \pi l_p)^6  Q_2 \delta^4(\vec y) \delta^4(\vec r)
\label{harm} \ , \ee
which can be inferred most efficiently from
\be d *_{11} G_4 = {1 \over 2} G_4 \wedge G_4 + (2 \pi l_p)^6  Q_2 \delta^4(\vec y)
\delta^4(\vec r)\, d^4 \vec y \wedge d^4 \vec r \,  .\label{gwedgeg}\ee
Here we allowed some arbitrary source of M2-brane charge sitting at
the origin, with a magnitude $Q_2$ that will be determined below.

Upon reduction to type IIA, equation \eqref{harm} reads
\be 0 =
\left(1 + {kR_{11} \over 2 r}\right)^{-1} \left( {\partial^2 \over \partial r^2} + {2 \over
r} {\partial \over \partial r}\right) H(y,r)+
\nabla^2_y H(y,r) + {l^2 V^4\over 2  r^4}  \delta^4(\vec y) + (2 \pi l_s)^5 g_s  Q_2
\delta^4(\vec y) \delta^3(\vec r)
\label{harm2} \ . \ee
In type IIA string theory we interpret the source as coming from
D2-branes, D4-branes and D6-branes sitting at the origin, so we have
\be Q_2 =   N_2 + b_0 N_4 + {N_6
b_0^2 \over 4} \,, \label{branesource} \ee
where $b_0$ is the value of the $b$ field at the position of the
branes, and the source arises from the presence of the Chern-Simons
coupling $e^{F+B} \wedge C$ on the worldvolume of the D-branes,
where $C$ is the sum of the RR potentials $A_i$ (we assume that no
field strengths are turned on in the D-brane worldvolumes).
Assuming that $H(r,y)$ is solved with respect to appropriate
sources, and that the parameters $\alpha$ and $l$ are set to
appropriate values, this background, in the suitable decoupling
limit, is dual to the $U(N_2 + N_4) \times U(N_2)$ quiver gauge
theory with $N_6$ fundamental matter multiplets.\footnote{Since
$2+1$ dimensional gauge theories are always asymptotically free, the
dual gravitational backgrounds are always highly curved in the UV
region (large radial coordinate).  Our discussion of these
backgrounds will be limited to smaller values of the radial
coordinate, where the solutions are weakly curved.} These solutions
are $2+1$ dimensional versions of a construction presented in
\cite{Grana:2001xn,Polchinski:2000mx}.  In fact, precisely this background was
constructed also in \cite{DiVecchia:2001uc}, with one important
caveat:
\cite{DiVecchia:2001uc} considered only the case where
\be b_\infty \equiv b(r \rightarrow \infty) = {1 \over 2}. \ee
As we will see shortly, allowing $b_\infty$ to take other values
provides opportunities to explore many interesting physical features
of this supergravity background.

\subsection{Charge quantization}

To construct the gravitational background dual to the $U(N_2+N_4)
\times U(N_2)$ gauge theory we still need to fix the values of the
parameters $\alpha$ and $l$. These parameters, which characterize
the four form $G_4^{SD}$, are naturally related to the number of
fractional branes $N_4$, and to the value of the $b$ field at
infinity, which is related \cite{Klebanov:1998hh,Morrison:1998cs} to
the ratio between the gauge couplings of the two groups. In fact, we
see immediately from the fact that $V(r) \rightarrow 1 $ as $r$ is
taken to be large, that
\be l = -2 \pi^2 k l_s^2 R  b_\infty - \alpha \ . \label{lrel} \ee
All that remains is to extract one more linear relation relating $l$,
$\alpha$, and $N_4$.

However, there is one small complication in specifying these
parameters. Recall that the gauge-invariant 4-form field strength in
type IIA supergravity is given by
\be \tilde {F}_4 = d A_3 + d B_2 \wedge  A_1 \ . \label{giffour} \ee
Since the D4-branes are wrapped on $\Sigma$, their flux will thread
${\cal M} \times S^2$ where $S^2$ is the sphere surrounding the
D6-branes. Using our solution above, the gauge-invariant flux through
${\cal M} \times S^2$ at some fixed $r$ is then
\be \int_{{\cal M} \times S^2} (-\tilde F_4) =  -2 \pi l V(r),  \label{max} \ee
which depends continuously on $r$.

On one hand, this is directly analogous to the continuous evolution of
the charge in the Klebanov-Strassler construction
\cite{Klebanov:2000hb}, which was interpreted as a cascade of
dualities.  It is natural to expect that the presence of fractional
branes in the $2+1$ dimensional context may give rise to some similar
duality cascade.  On the other hand, it is peculiar for charges, which
are generally quantized and conserved, to vary continuously as the
Gaussian surface surrounding the sources is varied.

An elegant resolution to this apparent conflict is reviewed by Marolf
in \cite{Marolf:2000cb}. There are three similar, yet distinct,
notions of charges: Maxwell charge, Page charge, and brane charge. We
will illustrate the difference between these charges in the example of
the D4-brane charge in type IIA string theory which was mentioned
above.

Maxwell charge is the flux of a gauge-invariant field strength (such
as (\ref{giffour})), measured at infinity; in the example above it is
the $r\to \infty$ limit of \eqref{max}. The corresponding current is
defined by
\be - d{\tilde F}_4 = *j_5^{Maxwell}. \ee
From the definition it is clear that this charge is gauge-invariant
and conserved. However, in general it is not quantized, and in the
presence of Chern-Simons-type terms in the bulk, it obtains
contributions from these terms as well as from the localized sources;
recall that in the absence of sources the Bianchi identity for
${\tilde F}_4$ takes the form $d{\tilde F}_4 = - F_2 \wedge H_3$.
This is what allows this charge to continuously depend on the radial
coordinate, as in \eqref{max}.

Brane charge, on the other hand, is the charge coming purely from the
localized sources, such as (\ref{branesource}) in the case of D2-brane
charge. In the D4-brane example it is defined by
\be - d{\tilde F}_4 -
F_2 \wedge H_3 = *j_5^{brane}. \label{bianchi} \ee
The definition implies that it is gauge-invariant, but in general it
is not conserved or quantized. The localized sources contributing to
the brane charge can come from various branes that carry the
appropriate charge. In our example it takes the form $\ast
j^{brane}_5=\ast j_5^{D4} + B_2\wedge \ast j_7^{D6}$, where the first
term is the D4-brane source current, including contributions from
gauge fields on D6-branes $F_2^{D6} \wedge *j_7^{D6}$ which count
D4-branes in bound states with D6-branes, while the second term is a
D4-brane current induced on D6-branes in the presence of a $B_2$
field.  These brane currents are normalized such that for a D-brane
spanning the directions $x_0 \ldots x_p$,
\be j_{p+1}^{Dp} = (2 \pi l_s)^{7-p} g_s\delta^{9-p}(\vec{y})\,  dx_0 \wedge dx_1 \ldots \wedge dx_p \ . \ee

We expect that a quantized charge should just measure the (integer)
number of D4-branes, and the correct charge that does this in the
presence of Chern-Simons terms (and respects Gauss' law, in that it
has the same value for any surface surrounding the sources) is the
Page charge.  The Bianchi identity (\ref{bianchi}) can be rewritten in
the form
\be  d (- \tilde F_4 -  F_2 \wedge B_2) = \ast j_5^{D4}\equiv \ast j_5^{Page}\ .
\label{pagefour}
\ee
Note that the induced D4-brane current from the D6-branes in the
presence of a $B_2$ field is canceled due to the D6-brane Bianchi
identity\footnote{In the presence of NS5-brane sources, there is an
induced D4-brane current $A_1\wedge \ast j^{NS5}_6$ on the NS5-brane
sources. This is canceled against the term $A_1\wedge H_3$ in
$\tilde{F}_4$ due to the NS5-brane Bianchi identity $dH_3 = \ast
j^{NS5}_6$.} $dF_2=\ast j_7^{D6}$; however, in the presence of gauge
field flux on the D6-branes they do still contribute to the D4-brane
Page charge. Equation \eqref{pagefour} implies that the D4-brane Page
charge, which is the integral of
\be -\hat F_4 \equiv -\tilde F_4 - F_2 \wedge B_2 =
- d (A_3 + B_2 \wedge A_1), \label{d4page} \ee
will respect Gauss' law and count the number of D4-branes, at the
expense of not being gauge-invariant (under gauge transformations of
the $B_2$ field).  This expression for the integer quantized flux can
also be inferred from the consideration of charge quantization in the
D1-D3 system, reviewed in Appendix \ref{appA}.

The D4-brane Page charge in our example of the previous subsection is
the flux of $\hat F_4$ on ${\cal M} \times S^2$, given by
\be \int_{{\cal M} \times S^2} (-\tilde F_4 - F_2 \wedge B_2) =  2 \pi \alpha.\ee
This turns out to be independent of $r$, as expected, due to a
delicate cancelation.  The Page charge is generally quantized,
implying in our case the quantization condition
\be  2 \pi \alpha = (2 \pi l_s)^3 g_s  N_4. \label{arel}\ee
Thus, we have identified the values of the undetermined parameters
$\alpha$ and $l$ in terms of the field theory data $N_4$ and
$b_\infty$ (we will review how $b_{\infty}$ is related to the gauge
couplings below).

To summarize, there are three distinct types of charges: Maxwell,
Page, and brane. These charges have different characteristics with
regards to the properties of localization, quantization, gauge
invariance, and conservation, which we summarize in table
\ref{table1}. Localization refers to the fact that the sources are
delta-function supported, and that the fluxes satisfy Gauss' law.
Quantization refers to the fact that charges are integers, in
appropriate units.\footnote{Up to certain overall shifts which we will
describe in more detail later.}  Gauge-invariance refers to invariance
with respect to all gauge transformations. Conservation refers to the
property that the total charge is invariant under dynamical evolution,
such as time evolution or moving along a moduli space. In simple
situations, all three notions of the charge coincide, so that all four
properties can be satisfied simultaneously. In the presence of
Chern-Simons terms and various background fluxes, however, the charges
are distinct.

\begin{table}
\centerline{\begin{tabular}{|l|c|c|c|c|c|} \hline
 & Localized  & Gauge-Invariant & Quantized & Conserved \\ \hline
Brane Charge & Yes & Yes &  & \\ \hline
Maxwell Charge &  & Yes & & Yes \\ \hline
Page Charge & Yes & & Yes & Yes \\ \hline
\end{tabular}}
\caption{\label{table1} The properties of different types of charges. The properties
not marked by ``Yes'' hold only in some cases.}
\end{table}

The fact that the Page charge that we computed above is not
gauge-invariant might raise concern as to whether this quantity can be
given any physical meaning. Indeed, non-gauge-invariant quantities are
always unphysical. In the case at hand, however, the gauge ambiguity
of the Page charge is only due to large gauge transformations which
shift $B_2$ by an element of the integer cohomology $H^2({\bf Z})$.
As long as we consider the Page charge modulo such a large shift, its
value is physical.  Since at infinity
\be -{1 \over (2 \pi l_s)^3 g_s} \int_{{\cal M} \times S^2} F_2 \wedge B_2 = - {k \over 2}
b_\infty, \ee
the large gauge transformation $b \rightarrow b+1$ causes $N_4$ to be
ambiguous modulo $k/2$. Note that for odd $k$ this suggests that the
quantization of $N_4$ is in units of $1/2$; we will elaborate more on
this issue later in this article.  Ignoring the subtlety, for the time
being, we learn that the ambiguity in $N_4$ is modulo discrete
shifts. Note that these shifts also change the value of the D2-brane
charge at the same time, as we will describe in more detail below.

The ambiguity in the Page charge has a beautiful physical
interpretation.  It means that the backgrounds dual to field theories
which lead to different Page charges, related by the large gauge
transformation, really have the same charges, so different field
theories could lead to the same gravitational background, and one
could have ``cascading'' RG flows between these theories (without
needing any brane sources in the bulk).  In the corresponding gravity
solutions, the $b$ field (and the Maxwell charge) changes continuously
as a function of the radial coordinate. It is often most convenient to
describe the solution in terms of $0 \leq b < 1$, and in order to do
this in some range of the radial coordinate one sometimes needs to do
large gauge transformations of $b$. These change the Page charge, and
thus also the field theory which we use to describe the solution in
the corresponding range of energies.  Various dual descriptions have a
window of energy scales where they are the most effective description
of the system.\footnote{A similar structure appears in the context of
Morita-equivalences of non-commutative field theory
\cite{Hashimoto:1999yj}.}  In some cases, such as
\cite{Klebanov:2000hb}, these flows are related to Seiberg
duality\cite{Seiberg:1994pq}; the interpretation of this via Page
charges was described in \cite{Benini:2007gx}. In other cases
\cite{Grana:2001xn,Polchinski:2000mx,Bertolini:2000dk,Bertolini:2001qa,
Bertolini:2001ma,Aharony:2000pp}, like the solution we wrote above,
the RG flows represent Higgsing of one gauge group to another.

Note that in the $3+1$ dimensional case, the field theories involved
in cascades are not asymptotically free, so the cascade always had to
be cut off somehow at large energies (large radial coordinates),
otherwise it leads to gauge groups of arbitrarily large rank at high
energies. In $2+1$ dimensions any gauge theory is asymptotically free,
so a key feature of cascades in $2+1$ dimensions is that the gauge
couplings are expected to asymptote to fixed values in the UV
\cite{Witten:1997sc}. There are no obstructions such as ``duality
walls'' which prevent one from flowing all the way to the UV
\cite{Benini:2007gx}. This is what makes parameters such as $b_\infty$
(related to the gauge couplings) and the Maxwell charges well-defined
in our construction above.

In the solution we wrote above, using the careful identification of
the parameters $\alpha$ and $l$ in terms of $N_4$ and $b_\infty$, we
can write
\be  b(r)  = b_\infty V(r) - {2 N_4 \over k} (1-V(r)), \ee
where $V(r \rightarrow \infty) = 1$ and $V(r\rightarrow 0) = 0$.  This
demonstrates explicitly how $b(r)$ interpolates from a continuous
value $b_\infty$ in the UV, to a discrete value $(- {2 N_4/ k})$ in
the IR.\footnote{ Note that parameters similar to $b_\infty$,
describing the positions of the D5-branes in the defect field theory,
should exist in the complete description of M-theory on $TN_{k1}
\times TN_{k2}$ mentioned in footnote \ref{fn} above, although these
parameters are decoupled in the limit where one flows to the $2+1$
dimensional theory dual to $ALE_{k1} \times TN_{k2}$.}

\subsection{D4-brane probes and the gauge couplings}
\label{dfour_probe}

One way to determine the relation between the string coupling $g_s$,
$b_{\infty}$, and the gauge couplings $g_{YM}^2$ of the two groups of
the field theory dual, as well as to determine the range of the
effective window for a given duality frame in a cascade, is to analyze
probe D4-branes and anti-D4-branes extended along ${\bf R}^{1,2}$ and
wrapping $\Sigma$ in this geometry.  Recall that we expect to have a
$U(N_2+N_4)\times U(N_2)$ ${\cal N}=4$ supersymmetric gauge theory,
which arises by splitting the $N_2$ ``regular D2-branes'' into two
types of ``fractional D2-branes,'' which can be thought of as a
D4-brane and an anti-D4-brane wrapping $\Sigma$, such that the total
D4-brane charge cancels, but such that the two branes together carry
one unit of D2-brane charge (so, if one of them carries $n$ units of
D2-brane charge, the other should carry $(1-n)$). Our original
description involved D4-branes that carry no D2-brane charge, so it
corresponds to $n=0$.  These gauge theories have a Coulomb branch
corresponding to separating all these ``fractional branes'' along the
directions transverse to the ALE space. On the gravity side, moving
along this Coulomb branch is realized by moving around fractional
branes in the background.

The general form of the D4-brane action is
\be S_{D4}   =  S_{DBI}+S_{WZ},\ee
with
\beq S_{DBI} & = & - T_4   \int_{R^{1,2} \times \Sigma} d^5 x\  e^{-\phi}
\sqrt{-\det(G+F+B)} ,  \\
S_{WZ} & = & - T_4 \int_{R^{1,2} \times \Sigma} \left[ {1 \over 2}(F+B) \wedge (F+B) \wedge
A_1 + (F+B) \wedge A_3 + A_5 \right] ,\eeq
and
\be T_p\equiv {1 \over g_s} (2\pi)^{-p} l_s^{-(p+1)}\ . \ee
We decompose
\be S_{DBI} = - T_4 \int_{R^{1,2} \times \Sigma} d^5 x\ e^{-\phi} \sqrt{-\det_{R^{1,2}}(G+B+F)}\sqrt{\det_{\Sigma}(G+B+F)},\ee
and proceed to extract the moduli space metric by allowing only the
time derivatives of the position of the D4-brane in the space
transverse to the D6-branes to be non-zero, in which case
\beq G_{00} &=& -H^{-1/2} V^{1/2}(1 + HV^{-1} (\dot r^2 + r^2 \dot \theta^2 + r^2\sin^2
\theta \dot \phi^2)), \cr
G_{11} & = & H^{-1/2} V^{1/2}, \qquad
G_{22} = H^{-1/2} V^{1/2}.
\eeq
We also take a vanishing worldvolume gauge field in the field theory
directions, $F_{01} = F_{02} = F_{12} = 0$, so that
\be -e^{-\phi} \sqrt{-\det_{R^{1,2}}(G+B+F)} = - H^{-1}\left[1
- {H V^{-1} \over 2}(\dot r^2 + r^2 \dot \theta^2 + r^2\sin^2 \theta \dot
\phi^2)+ \cdots\right].\ee
On the other hand, the contribution from the vanishing 2-cycle directions is
\be \int_{\Sigma} \sqrt{-\det_{\Sigma}(G+B+F)} = {2 \over k R_{11} } |
2 \pi^2 l_s^2 R_{11} k n \pm  (lV +\alpha)|, \ee
where $\pm$ distinguishes a D4-brane from an anti-D4-brane, and $n$ is
the quantized magnetic flux of the worldvolume gauge field $F$ through
$\Sigma$ (which is related to the D2-brane charge of the probe). The
contribution from the Chern-Simons terms is
\be S_{WZ} =  T_4 \int {2 \over k R_{11}} \left[2 \pi^2 l_s^2 R_{11} k n  \pm  (lV
+\alpha)\right] H^{-1}+ \cdots. \ee
We can consider probes carrying different amounts of D2-brane charge;
usually we will only look at the lightest D4-brane and anti-D4-brane,
and view branes with other charges as bound states of these lightest
branes with D2-branes.

Now, we see that
\be S = S_{DBI}+ S_{WZ} \ee
simplifies drastically provided
\be 2 \pi^2 l_s^2 R_{11} k n  \pm  (lV +\alpha) > 0. \label{bound} \ee
In this case the terms with no derivatives cancel, and the relevant terms in
the action have
the form
\beq S &=& T_4 \left( \left(2 \pi^2 l_s^2 n \pm {\alpha \over kR_{11}}\right)V^{-1}\pm {l
\over kR_{11}} \right)(\dot r^2 + r^2 \dot \theta^2 + r^2\sin^2 \theta \dot \phi^2)  \cr
& \equiv & {2 \pi^2 l_s^2 T_4 \over g_{eff}^2(r)} (\dot r^2 + r^2 \dot \theta^2 + r^2\sin^2
\theta \dot \phi^2),   \eeq
where in the last line we defined a dimensionless effective coupling
$g_{eff}^2(r)$ on the worldvolume (note that this coupling also
appears in the kinetic term of the worldvolume gauge field).
Substituting $l$ and $\alpha$ in terms of $b_\infty$ and $N_4$, using
the results (\ref{lrel}) and (\ref{arel}) from the previous
subsection, we have
\be {1 \over  g_{eff}^2(r)} =  (n \pm b_\infty) + {R_{11} (nk \mp 2 N_4 ) \over 2 r} \ . \ee
Finally, we perform the standard map between gauge theory and dual
string theory parameters
\be R_{11} = g_s l_s, \qquad r = 2 \pi l_s^2 \Phi, \qquad g_s = g_{YM}^2 (2 \pi)^{-(p-2)}
l_s^{-(p-3)}, \ee
where $\Phi$ is the vacuum expectation value of the scalar field in
the ${\cal N}=4$ gauge theory, to find
\be
{1 \over  g_{eff}^2(\Phi)} = (n \pm b_\infty) + {g_{YM}^2 (nk \mp 2 N_4) \over 4 \pi \Phi} \
. \ee
This should be interpreted as the running coupling on the moduli space
of the field theory, as we move along the Coulomb branch.

If one takes, for concreteness, the naive fractional brane values that
we mentioned above, $n = 0$ for the ``+'' (D4-brane) and $n=1$ for the
``$-$'' (anti-D4-brane), the effective gauge couplings take the form
\beq
{1 \over  g_{eff1}^2(\Phi)} &=& b_\infty  - { g_{YM}^2 N_4 \over 2 \pi \Phi} \ , \\
{1 \over  g_{eff2}^2(\Phi)} &=& (1 - b_\infty) + {g_{YM}^2(k + 2 N_4) \over 4 \pi \Phi} \ .
\eeq
We expect these to be the inverse effective couplings on the moduli
space of our ${\cal N}=4$ $U(N_2+N_4)\times U(N_2)$ gauge theory (up
to an overall factor of $g_{YM}^2$), and this is indeed the
case. Recall \cite{Seiberg:1996nz} that in $d=3$ ${\cal N}=4$ gauge
theories, the effective gauge coupling as a function of position in
the moduli space is one-loop exact.  There is a constant term which is
just the classical gauge coupling, and a one-loop correction,
independent of the Yang-Mills coupling, proportional to the number of
hypermultiplets in the fundamental representation $N_f$ minus twice
the number of colors $N_c$, divided by the radial position $\Phi$ on
the Coulomb branch. This matches with the equations above, if we
identify the gauge couplings of the two groups as $1/g_1^2 =
b_{\infty}/g_{YM}^2$ and $1/g_2^2 = (1-b_{\infty})/g_{YM}^2$; these
are precisely the expected couplings for D4-branes wrapping a
vanishing cycle with a $B_2$ field proportional to $b_{\infty}$. The
number of flavors of the $U(N_2+N_4)$ group is $2N_2$ (from the two
bifundamentals), so for this group $N_f-2N_c = -2N_4$, while the
$U(N_2)$ gauge group has $2(N_2+N_4)+k$ fundamental hypermultiplets,
so it has $N_f-2N_c = 2N_4+k$. This agrees with the expressions above
when we choose the $k$ fundamental hypermultiplets to be charged under
the $U(N_2)$ group.

\begin{figure}[t]
\hspace{-0.25in}\begin{tabular}{cc}
\hspace{-0.5in} \includegraphics[width=4in]{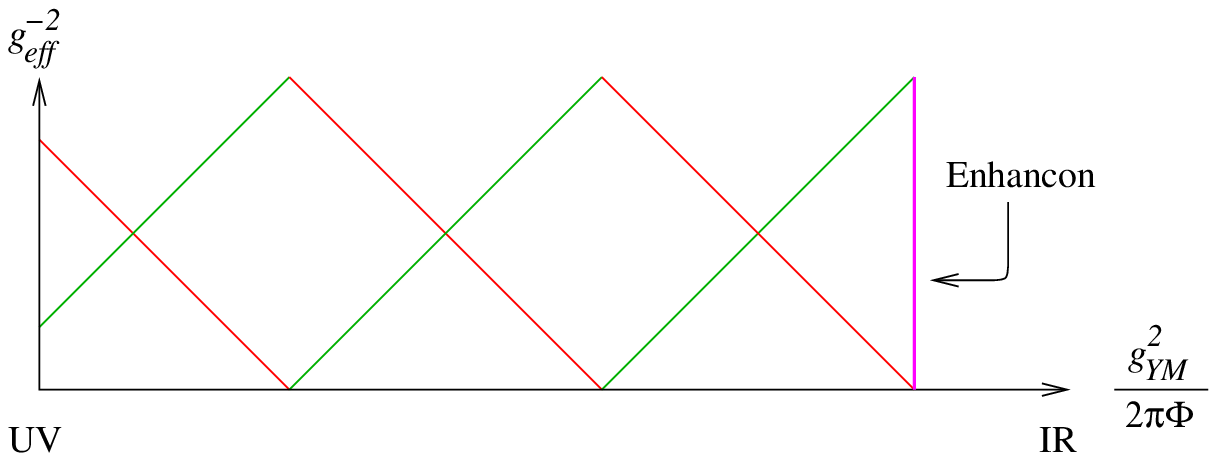}
\hspace{-0.5in}
&
\hspace{-0.5in}
\includegraphics[width=4in]{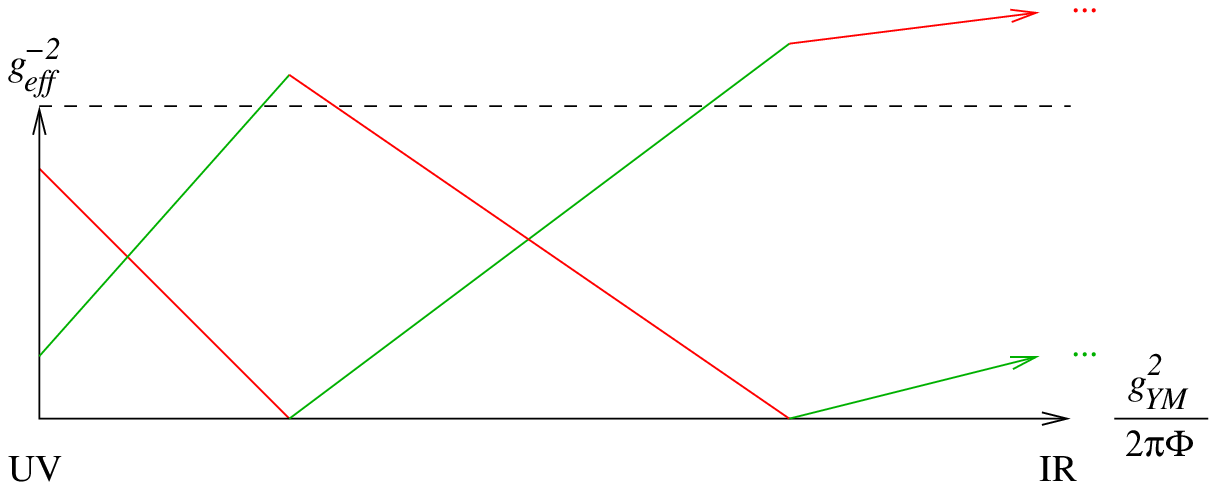}
\hspace{-0.5in} \\
(a) & (b) \end{tabular}
\caption{\label{figa} The effective gauge couplings of the two gauge
groups on the moduli space, as a function of the moduli space
coordinate; in theories with eight supercharges this may also be
interpreted as the effective running coupling at a particular
scale. Figure (a) represents the $k=0$ case where the cascade
terminates in the IR with an enhancon.  Figure (b) represents a
cascade ending when $k > 2 N_4$, so that the effective gauge couplings
become small as $\Phi \to 0$ (and there may be a non-trivial SCFT
there).}
\end{figure}

As one moves the D4-branes along the Coulomb branch, there is a point,
\be \Phi = {g_{YM}^2 N_4 \over  2 \pi  b_{\infty}}, \ee
at which the effective coupling $g_{eff1}^2$ becomes infinite. This is
an enhancon point \cite{Johnson:1999qt}, where the tension of the
D4-brane wrapped on $\Sigma$ is going to zero. Generally, in
backgrounds like this, there are tensionless D4-branes sitting at the
enhancon in the background
\cite{Polchinski:2000mx,Bertolini:2000dk,Aharony:2000pp}, although
they did not appear explicitly in our solution. If we try to move the
D4-brane beyond this point, it no longer satisfies \eqref{bound}, so
we are no longer on the moduli space. However, there are different
probes that one can use beyond this point : we can use $n=1$ for the
``$+$'' (D4-brane) probe and $n=0$ for the ``$-$'' (anti-D4-brane)
probe.  For these probes, the effective couplings are
\beq
{1 \over  g_{eff1}^2(\Phi)} &=& (1+b_\infty) - { g_{YM}^2 (2 N_4 - k)\over 4 \pi \Phi} \
,\\
{1 \over  g_{eff2}^2(\Phi)} &=& (1 - (1+b_\infty)) + {g_{YM}^2 N_4 \over 2 \pi \Phi} \
,\eeq
which are still positive in this region.  Note that for $k=0$ the
change we made is the same as shifting $b_\infty$ by one; the gauge
transformation that shifts $b_{\infty}$ by one also shifts $n$, and
only their sum (for a D4-brane) is gauge-invariant.  We will provide
further interpretation of this shift in section \ref{brane_nfour}
below, but for now, simply note that we can interpret the change in
the effective action on the moduli space as related to a RG flow along
a cascade (since, for theories with eight supercharges, the running
couplings are the same as the effective couplings on the moduli
space).  This is best illustrated by drawing the inverse couplings on
the two types of brane probes (chosen such that both probes have
positive couplings) as a function of $g_{YM}^2 N_4 /2 \pi \Phi$ (see
figure \ref{figa}).
Note that with each cascade transition, $N_2$ decreases (we will
describe the precise change in more detail below).  The reduced rank
of the gauge group means that we have a smaller moduli space; the
coordinates of the moduli space which ``disappeared'' correspond to
the positions of the D4-branes sitting at the enhancon point (see
\cite{Benini:2008ir} for details).

The cascade can end in the IR in one of two ways (see figure
\ref{figa}). One possibility is that we encounter an enhancon where
all the branes sit (when this enhancon is at $\Phi > 0$, the solution
inside the enhancon radius is just the solution with no fluxes). This
is the only possibility when $k=0$, and it also happens whenever no
further transitions can be made (with positive rank gauge groups).  In
this case one expects the enhancon mechanism to capture the IR
dynamics of the theory, namely the background should contain
fractional branes sitting at the enhancon carrying all the remaining
charges. The second possibility, when $k>0$, is that the value of
$N_4$ could decrease until $N_4<k/2$, and then both couplings start
decreasing towards the IR so there are no further cascades.

\subsection{The harmonic function and the decoupling limit}

\begin{figure}
\centerline{\includegraphics[width=3in]{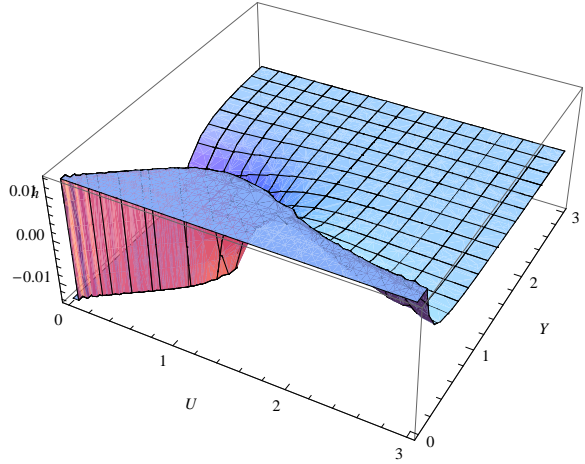}}
\caption{\label{figb} The solution to the harmonic equation in the decoupling
limit, for $k=0$.}
\end{figure}

For the sake of completeness, let us briefly discuss the solution to
the harmonic equation (\ref{harm2}) and the decoupling limit. If one
takes all of the D2, D4, and D6-branes to be coincident at the origin,
(\ref{harm2}) takes the form
\be - V \left( {\partial^2 \over \partial r^2}+  {2 \over r} {\partial \over \partial
r}\right) H(y,r)  - \nabla^2_y H(y,r) = {l^2 V^4\over 2  r^4}  \delta^4({\vec y}) + (2 \pi l_s)^5
g_s  Q_2 \delta^4({\vec y}) \delta^3({\vec r}).
\ee
In taking the decoupling limit, one performs the standard scaling
\be r = \alpha' U, \qquad y = \alpha' Y, \qquad g_{YM2}^2 = g_s l_s^{-1},  \ee%
and one should also rescale
\be H(r,y) = {1 \over \alpha'^2} h(U,Y)\ . \ee
In this limit,
\be
V^{-1} =  1 +  {g_{YM}^2  k \over 2 U}
\ee
is independent of $\alpha'$. We also scale $l$ and $\alpha$ according
to (\ref{lrel}) and (\ref{arel}).  In terms of these variables, the
harmonic equation takes the form
\beq  \lefteqn{V \left( {\partial^2 \over \partial U^2} + {2 \over U} {\partial \over
\partial U}\right) h(Y,U) + \nabla^2_Y h(Y,U)} \\
& =& -{(2 \pi)^4 g_{YM2}^2 \left(N_4 + {b_\infty k \over 2}\right)^2 V^4\over 2  U^4}
\delta^4({\vec Y})   - (2 \pi)^5 g_{YM2}^2 \left(N_2 + b_0 N_4 + {N_6 b_0^2 \over 4}\right)
\delta^4({\vec Y}) \delta^3({\vec U}),\nonumber
\eeq
and is independent of $\alpha'$. Just as in \cite{Cherkis:2002ir}, all
of the structures of (\ref{harm2}) survive in this scaling limit.  In
terms of the solution to this warp factor equation, the supergravity
background metric can be written in the standard form
\be {ds_{IIA}^2 \over \alpha'} =  h^{-1/2} (-dt^2 + dx_1^2+dx_2^2) + h^{1/2}  dS_{ALE}^2 +
h^{1/2}  (dU^2 + U^2 (d \theta^2 + \sin^2 \theta d \phi^2)), \ee
where
\be d s_{ALE}^2 = \alpha'^2 dS_{ALE}^2 = \alpha'(d Y^2 + Y^2 ds_{Lens}^2) \ . \ee
Although somewhat cumbersome, $h(U,Y)$ can be determined using various
methods. As an example, we illustrate the form of $h(U,Y)$ for the
case of $k=0$ in figure \ref{figb}.

\subsection{Maxwell charges and the brane creation effect}
\label{brane_nfour}

Although we have identified the Page charge as the relevant quantity
for imposing charge quantization conditions, Maxwell charges remain an
important gauge-invariant parameter to distinguish between physically
inequivalent configurations.

As reviewed above, Maxwell charges are defined as the flux of the
gauge-invariant field strength at infinity.  In many cases, the
easiest way to compute the Maxwell charge is to relate it to the flux
defining the Page charge.  In particular, the fluxes of
\beq (-\hat F_4) &=& -\tilde F_4 - B_2 \wedge F_2, \\
\hat F_6 & = & * \tilde F_4 - B_2 \wedge (-\hat F_4) -{1 \over 2} B_2 \wedge B_2 \wedge F_2,
\eeq
define the Page charges of D4-branes and D2-branes, respectively, in
the type IIA theory.  This can then be used to compute the Maxwell
charges where we write
\beq Q_4^{Maxwell}&=&{1 \over (2 \pi l_s)^3 g_s}\int_{{\cal M} \times S^2} (-\tilde F_4) =
{1 \over (2 \pi l_s)^3 g_s} \int_{{\cal M} \times S^2} \left[(-\hat F_4) + B_2 \wedge F_2\right] ,
\label{q4max}\\
 Q_2^{Maxwell} &=& {1 \over (2 \pi l_s)^5 g_s}\int_{ALE \times S^2}  * \tilde F_4 \cr
&  = &
{1 \over (2 \pi l_s)^5 g_s} \int_{ALE \times S^2}  \left[\hat F_6 + B_2 \wedge (-\hat F_4) + {1
\over 2} B_2 \wedge B_2 \wedge F_2 \right] . \label{q2max}
\eeq
Evaluating these integrals at infinity, and denoting the quantized
Page charges by $N_4$ and $N_2$ as above, we obtain
\beq
Q_4^{Maxwell} &=& N_4 + {1 \over 2} b_\infty N_6, \label{qfourmax}\\
Q_2^{Maxwell} & = & N_2 + b_\infty N_4 + {1 \over 4} b_\infty^2 N_6. \label{qtwomax}\eeq
These gauge-invariant charges at infinity are well-defined in 2+1
dimensions. (In 3+1 dimensional cascading backgrounds, they grow
logarithmically and are therefore infinite.)

There are two qualitatively distinct ways in which the value of
$b_\infty$ can change. One is to vary it continuously, keeping the
Page charges $N_2$, $N_4$, and $N_6$ fixed. This is a physical
deformation, and it changes the values of $Q_2^{Maxwell}$ and
$Q_4^{Maxwell}$ accordingly. We refer to this operation as ``sliding''
$b_\infty$. Another possibility is to shift the value of $b_\infty$ by
one via a large gauge transformation. This also changes the values of
$N_2$ and $N_4$ (as well as transforming the background fluxes), but
since the gauge transformation is not a physical process, the
gauge-invariant $Q_2^{Maxwell}$ and $Q_4^{Maxwell}$ are unchanged. We
will refer to this procedure as ``shifting'' $b_\infty$. The ``shift''
can only vary $b_\infty$ by an integer amount. It is clear from the
equations above, that ``shifting'' $b_\infty \to b_\infty - 1$ takes
$N_4 \to N_4 + {1\over 2} N_6$ and $N_2 \to N_2 + N_4 + {1\over 4}
N_6$.

\begin{figure}[t]
\centerline{\includegraphics{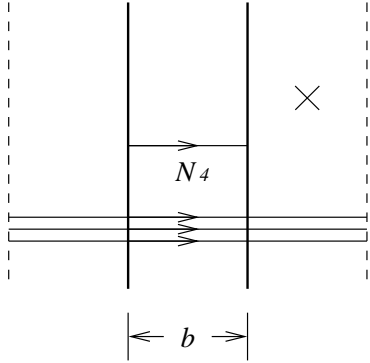}}
\caption{\label{figc} The brane configuration describing a
$U(N_2)\times U(N_2+N_4)$ theory with $N_6$ additional flavors charged
under $U(N_2)$. The vertical solid lines represent the NS5-branes
oriented along the 012345 directions.  The horizontal solid lines
represent the D3-branes extended along the 0126 directions. The
$\times$ represents the D5-brane extended along the 012789
directions. The vertical direction in the figure represents one of the
345 directions, whereas the horizontal direction in the figure
represents the $x^6$ coordinate. The vertical dashed lines in the
figure are identified as a result of the compactification along the
$x^6$ direction.  The $N_4$ fractional D3-branes stretch between the
NS5-branes, whereas $N_2$ integer D3-branes stretch around the
periodic direction.  Similar conventions apply to all of our brane
diagrams.}
\end{figure}

The behavior of Maxwell charges under combination of ``slides'' and
``shifts'' has interesting ramifications. It is useful to consider
this process from the point of view of the brane configurations which
reduce to the corresponding gauge theories at low energies
\cite{ori_branes,Hanany:1996ie}. These brane configurations involve
D3-branes stretched along a compact $x^6$ direction (as well as the
three field theory directions), two NS5-branes stretching along the
$x^3,x^4,x^5$ directions, and $N_6$ D5-branes stretching along the
$x^7,x^8,x^9$ directions. Given some value of $0<b_\infty<1$, our
claim is that the supergravity background (\ref{iiaf})--(\ref{iial})
is dual to the field theory described by the brane diagram illustrated
in figure \ref{figc}, where $b_{\infty}$ is the distance in $x^6$
between the two NS5-branes along the fractional branes (divided by the
radius of the $x^6$ circle). For the moment, we ignore the vertical
placement of the fractional branes and D5-branes in figure \ref{figc}.

\begin{figure}[t]
\centerline{\includegraphics{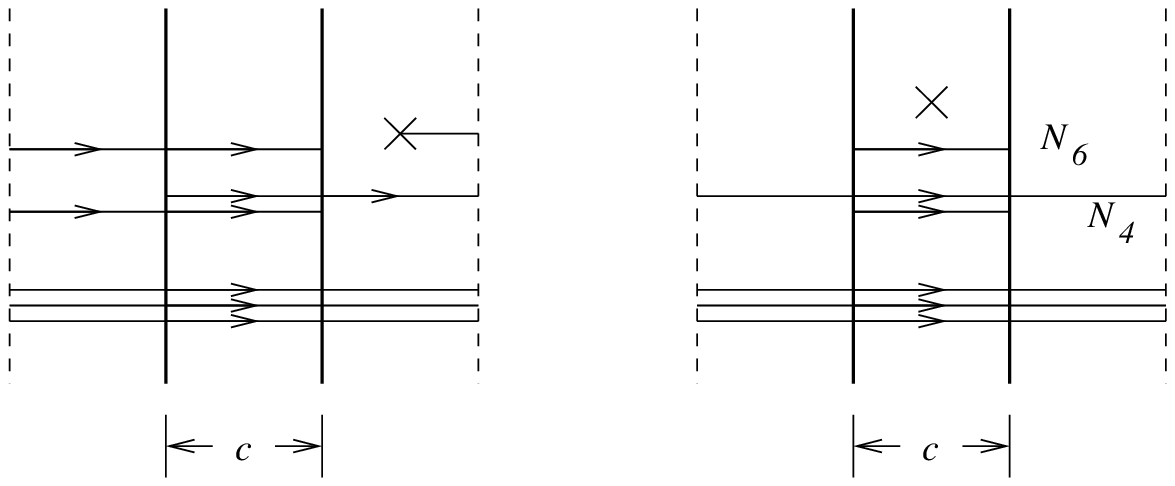}}
\caption{\label{figd} The same brane configuration after moving the
NS5-brane on the right around the circle to the right (on the left),
and (on the right) the same configuration after also moving the
D5-branes to the right.}
\centerline{\includegraphics{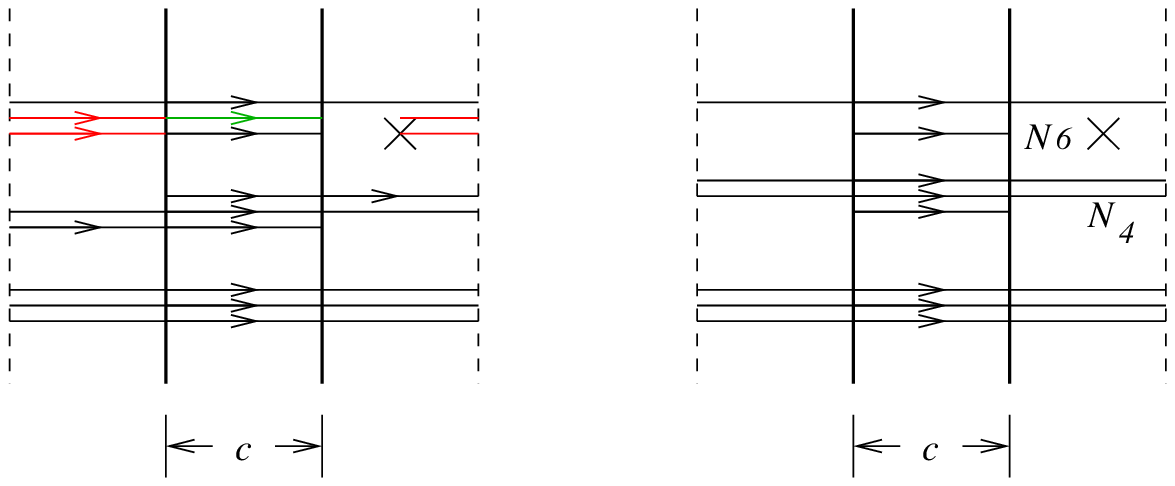}}
\caption{\label{fige} The brane configuration obtained after taking
an NS5-brane twice around the circle (on the left), and (on the right) the same
configuration after also moving
the D5-branes.}
\end{figure}

Suppose now we consider sliding $b_\infty$ so that we let $b_\infty =
1+c$ for some $0 < c < 1$. In the brane configuration we can realize
this by moving one of the NS5-branes around the circle. Taking into
account the brane creation effect when the NS5-branes cross the
D5-branes, the brane configuration becomes the one on the left-hand
side of figure \ref{figd}, with $N_6$ D3-branes ending on the
D5-branes. Now, taking advantage of the fact that moving the D5-branes
in $x^6$ while preserving (locally) the linking number does not change
the effective dynamics in 2+1 dimensions, we can move the D5-branes to
obtain a different description of the same low-energy theory on the
right of figure \ref{figd}.\footnote{It should be noted that as 3+1
dimensional defect theories, these two configurations are distinct,
but the difference decouples as one flows from 3+1 to 2+1, essentially
along the lines of the decoupling of $b_\infty$ in flowing to the IR
fixed point of the 2+1 dimensional theory.}

The configuration on the right in figure \ref{figd} shows the
D5-branes giving rise to flavors charged under the gauge group with
larger rank, even though we started with the opposite situation in
figure \ref{figc}. We will elaborate on the exchange in gauge group
with respect to which the flavors are charged in the next subsection.
For now, simply note that sliding to $b_\infty = c+2$ for $0<c<1$ will
lead to the configuration illustrated in figure \ref{fige}, where the
flavor branes are back in their original position.\footnote{It might
appear at first sight that the left figure violates the $s$-rule
because two D3's (in red) stretch from a D5 to the NS5. However, since
these D3's terminate on {\it different} NS5's (in green) they do not
violate the $s$-rule. We will discuss this issue further in the
following section.}

\begin{figure}[t]
\centerline{\includegraphics{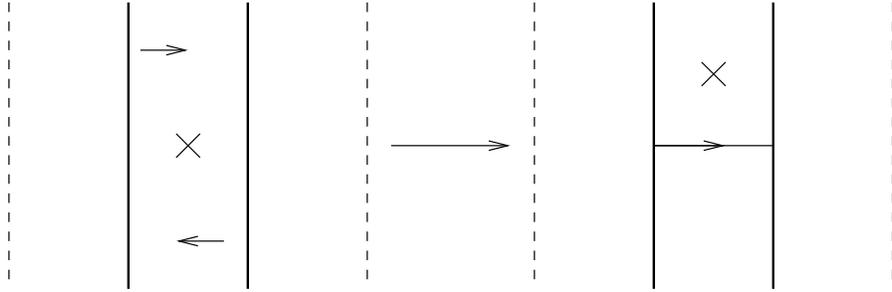}}
\caption{\label{figf} The ``basic transition'' that creates fractional
branes when two NS5-branes cross each other in the presence of a
D5-brane between them.}
\end{figure}

The changes in the number of fractional branes, as we slide
$b_\infty$, can be thought of as a consequence of a fundamental
process where an NS5-brane sandwiches a D5-brane as they cross one
another, as illustrated in figure \ref{figf}.

After one performs a slide so that $b_\infty=c+2$, it is useful to
perform a ``shift'' to bring the value of $b_\infty$ back to the value
$c$, obeying $0 < c < 1$. This is just twice the transformation
described earlier; since the ``shift'' keeps the Maxwell charges
intact, one can read off the variation in the Page charges by
rewriting
\beq
Q_2^{Maxwell} & = & N_2 + (c+2) N_4 + {1 \over 4} (c+2)^2 N_6 \cr
& = & (N_2 + 2 N_4 + N_6) + c (N_4 + N_6)  + {1 \over 4} c^2 N_6.\eeq

Note that the change we obtain in the Page charges (which arises also from
performing twice the single ``shift+slide'' transformation described above),
\beq N_2 &\rightarrow &N_2 + 2 N_4 + N_6, \\
N_4 & \rightarrow & N_4 + N_6,
\eeq
precisely matches the change in the counting of integer and fractional
branes illustrated in figure \ref{fige}. We therefore learn that the
precise forms of the Maxwell and Page charges carefully account for
the brane creation effects \cite{Hanany:1996ie}. In other words, one
can view the brane creation effects as the logical consequence of the
subtlety in the quantization of charges in the presence of
Chern-Simons terms in the low-energy effective description of M-theory
and type IIA supergravity. This will prove to be a powerful diagnostic
tool.

\subsection{Some generalizations}

We close this section by discussing a few generalizations of the
construction of the supergravity dual of ${\cal N}=4$ theories in
$2+1$ dimensions.

\subsubsection{Mass and moduli space deformation}

\begin{figure}
\centerline{\includegraphics{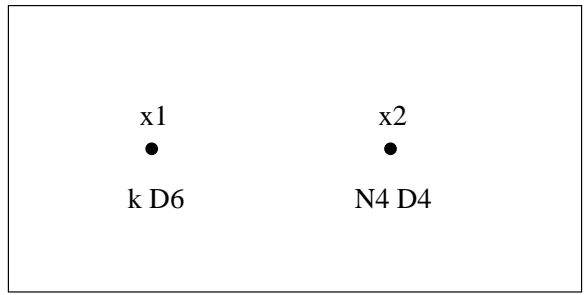}}
\caption{\label{figg} The positions in the transverse space
($x^7 \ldots x^9$) of the D4-branes and the D6-branes.}
\end{figure}

One simple generalization is to consider D2-branes, D4-branes, and
D6-branes placed at different points along the coordinates transverse
to the D6-branes.  Moving the D6-branes corresponds to changing the
masses of the fundamental matter fields, and was considered briefly in
\cite{Cherkis:2002ir}. Moving the D2-branes and the D4-branes
corresponds to moving along the Coulomb branch of the moduli space (as
we discussed for a single fractional brane in section
\ref{dfour_probe}). Here, for the sake of illustration, let us
consider separating all of the $N_4$ D4-branes away from the $k$
D6-branes, as illustrated in figure \ref{figg}.

It is not too difficult to verify that
\be G_4^{SD}= (2 \pi l_s)^2 R_{11}\,  \omega \wedge (\eta + *_4 \eta), \ee
for
\be
\eta = -\left({1 \over 2} b_\infty k V - N_4 V {R_{11} k \over 2| \vec x - \vec x_2|}
\right) d \Omega_1 - N_4 d \Omega_2 \ee
(where $d\Omega_1$ is the volume form of a 2-sphere around the
D6-branes, and $d\Omega_2$ is the volume form of a 2-sphere around
the D4-branes), is self-dual, closed, and has the appropriate
behavior at large distances (with appropriate sources). It should
not be too difficult to construct an analogous expression for a
general distribution of D4-branes and D6-branes in the three
transverse directions. As in \cite{Cherkis:2002ir}, solving for the
harmonic function $H$ is much more challenging in these general
cases.

\subsubsection{Flavors charged under $U(N_2)$ and $U(N_2+N_4)$ \label{s262}}

Another generalization we can consider is changing the gauge group
under which the fundamental hypermultiplets are charged. Recall that
our claim is that the supergravity background
(\ref{iiaf})--(\ref{iial}) is dual to the field theory described by
the brane configuration illustrated in figure \ref{figc}, with the
fundamental hypermultiplets charged under the lower rank group.  What
then would constitute the supergravity dual for the field theory
described by the brane configuration illustrated in figure \ref{figd},
with matter charged under the higher rank group ?

The answer to this question can be gleaned from the way in which we
obtain the configuration of figure \ref{figd} in the first place.  The
configuration of figure \ref{figd} is obtained by ``sliding''
$b_\infty$ by one, and then ``shifting'' it by $-1$ (if we want to
bring it back to $[0,1]$).  The brane configuration suggests that we
have a $U(N_2') \times U(N_2'+N_4')$ theory with
\be N_2' = N_2  + N_4, \qquad N_4' = N_4 + N_6. \ee
The Maxwell charge is the same as our original expression
\eqref{qfourmax},\eqref{qtwomax} for the charges when $b=b_\infty+1$
(since it is not changed by the ``shift''); expressing it in terms of
the new charges we obtain
\beq Q_2^{Maxwell} &=& \left(N_2' +{ N_6 \over 4} \right) + \left(N_4' - {N_6 \over
2}\right)b_\infty  + {N_6 \over 4} b_\infty^2, \\
Q_4^{Maxwell} & = & \left(N_4' - {N_6 \over 2}\right) + {N_6 \over 2} b_\infty.
\eeq
In terms of the new charges, the large gauge transformation has
shifted the Page charges to
\be Q_2^{Page} = N_2' + {1 \over 4} N_6, \qquad Q_4^{Page} = N_4' - {1 \over 2} N_6.
\label{newpage} \ee

The fact that this expression for the Page charges is different from
our previous expression is consistent with the fact that we are
describing a different configuration now, with the fundamental fields
charged under a different group. From the point of view of the type
IIA picture, a simple way to understand the charges \eqref{newpage} is
to say that we now have a self-dual gauge field flux on the
worldvolume of each of the D6-branes (using the normalization where
${\cal F} = B+2 \pi \alpha' F$),
\be  F = 2 \pi \omega_2 \ . \ee
Namely, we identify the D6-brane without the flux as giving a flavor
for one gauge group, and the D6-brane with this flux as giving a
flavor for the other gauge group.  A curious fact here is that these
self-dual fluxes are quantized such that the induced D4-brane charge
on the D6-brane worldvolume is half-integral.

\begin{figure}
\centerline{\includegraphics{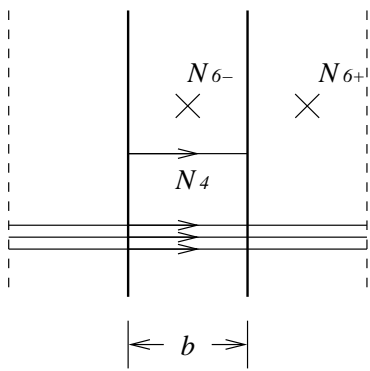}}
\caption{\label{figh} A brane configuration with flavors charged under
both gauge groups.}
\end{figure}

The problem with using the slide and shift of $b_\infty$ by one is
that one can only move all $k$ D5-branes in the type IIB description
at once under this transformation. What would it mean to move some of
the D5-branes, so that there are $N_{6-}$ flavors charged with respect
to the $U(N_2+N_4)$ gauge group and $N_{6+}$ flavors charged with
respect to the $U(N_2)$ group, as illustrated in figure \ref{figh}? If
we take the D6-branes in the type IIA description as probes, the
answer, based on the discussion above, is that we need to turn on a
self-dual flux for the $U(1)$ gauge field on the $N_{6-}$
D6-branes. This picture is consistent with what is described in
\cite{Grana:2001xn,Benini:2007gx} in related contexts.

As far as charges are concerned, we can write down expressions
consistent with our previous expressions by taking
\beq Q_2^{Maxwell} &=& \left(N_2 +{ N_{6-} \over 4} \right) + \left(N_4 - {N_{6-} \over
2}\right)b_\infty  + {N_6 \over 4} b_\infty^2, \label{maxwell1}\\
Q_4^{Maxwell} & = & \left(N_4 - {N_{6-} \over 2}\right) + {N_6 \over 2} b_\infty,
\label{maxwell2}
\eeq
and
\be Q_2^{Page} = N_2 + {1 \over 4} N_{6-}, \qquad Q_4^{Page} = N_4 - {1 \over 2} N_{6-},
\ee
where we have dropped the primes from the $N_2$'s and the $N_4$'s.
These Page charges can then be used to construct the supergravity
duals to these theories using (\ref{iiaf})--(\ref{iial}) and
(\ref{arel}) and (\ref{lrel}), and using the appropriate Page charges
to specify the discrete adjustable parameters of the solution.

There is one issue to keep in mind: while there are 4 sets of discrete
data, $N_2$, $N_4$, $N_{6+}$, and $N_{6-}$, in the field theory, there
are only three Page charges $Q_2^{Page}$, $Q_4^{Page}$, and
$Q_6^{Page}$ which determine our gravity solution.  Thus, there is a
degeneracy in the values of the Page charges corresponding to distinct
field theories.  This degeneracy is most likely lifted by taking into
account the fluxes on the D6-branes; in similar cases these can be
thought of as Wilson lines on the D6-branes
\cite{Gaiotto:2009tk,Hikida:2009tp,Fujita:2009xz}, which, when lifted to M theory,
become topological data of the 3-forms \cite{Witten:2009xu}. While
there are sufficient degrees of freedom in the topological data of the
3-forms to lift the degeneracy between the field theory and gravity
data, identifying the one-to-one map between gauge equivalence classes
of gravity backgrounds and distinct field theories seems rather
challenging. This issue is closely related to that of mapping out the
moduli space of the field theory from the supergravity point of view,
which remains generally a subtle issue. We leave detailed
consideration of these issues for future work.

\section{Flux quantization in the ABJM and ABJ models}

In this section, we repeat the analysis of quantization of charges and
fluxes for the ABJ and ABJM systems (with Yang-Mills kinetic terms),
namely for $U(N_2)_k\times U(N_2+N_4)_{-k}$ Yang-Mills-Chern-Simons
theories with ${\cal N}=3$ supersymmetry in three dimensions (for
simplicity, in this case we do not add any matter fields to those
already present in the ABJM construction).  Our goal is to demonstrate
the consistency of having a continuous value of $b_\infty$ but a
discrete value for $b$ in the IR, and to understand better how the
gauge potential flux $b$ is quantized as a consequence of charge
quantization. We will reverse the order of presentation of the
previous section, starting by looking at the counting of branes in the
corresponding brane configuration, and inferring from this an expected
form for the Maxwell and Page charges.  Although we will not
explicitly construct the dual supergravity solution here, we will have
enough information from charge quantization to reproduce similar
conclusions to those of section 2. We will then show that the
quantization of $b$ thus obtained is consistent with a variety of
checks.

\subsection{Brane configurations and Maxwell charges\label{s31}}

\begin{figure}
\centerline{\includegraphics{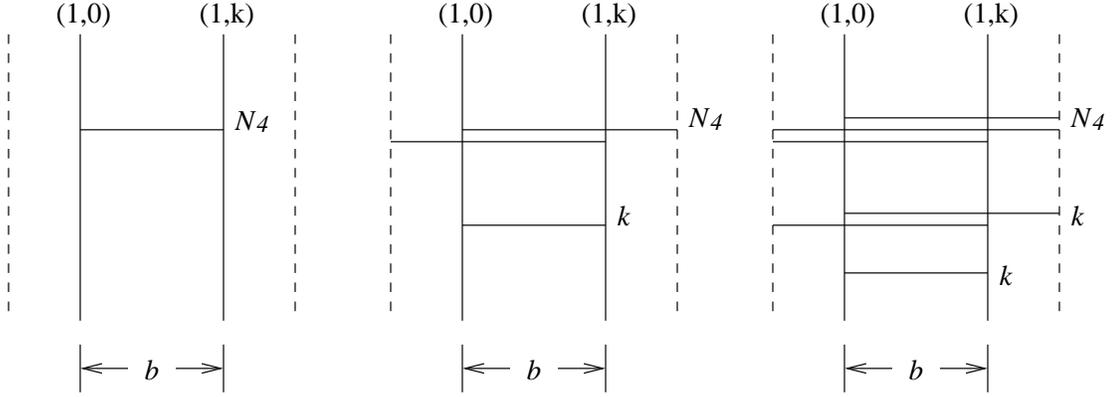}}
\caption{\label{figi}The brane configuration for the ${\cal N}=3$
theories, and its change upon ``sliding'' $b\to b+1 \to b+2$. In the
${\cal N}=3$ theories, fractional branes are {\it not} free to move in
the vertical direction.  We will nonetheless separate the branes in
the vertical directions to avoid cluttering the figure.  The $N_2$
integer branes, winding all the way around the periodic direction,
have also been suppressed in the figure.}
\end{figure}

The ${\cal N}=3$ Yang-Mills-Chern-Simons theories arise as the
low-energy effective field theory of D3-branes stretched on a circle,
intersecting one NS5-brane and one $(1,k)$ 5-brane in type IIB string
theory. Models of this type were originally considered in
\cite{Kitao:1998mf,Bergman:1999na}. Following the convention of
\cite{Aharony:2008ug}, we take the D3-branes to be oriented along the
coordinates 0126 (with $x^6$ on a circle), the NS5-brane to be
oriented along 012345, and the $(1,k)$ 5-brane to be oriented along
$012[3,7]_\theta [4,8]_\theta [5,9]_\theta$ with $\tan(\theta) = {1 /
g_s k}$ (assuming that the RR axion vanishes). We allow some number
$N_2$ of ``regular branes'' which are D3-branes going all the way
around the circle, and some number $N_4$ of ``fractional branes,''
which are D3-branes stretched just along one segment of the circle.
Such a brane configuration is illustrated in figure \ref{figi}. As in
the previous section, in the brane configuration we have a parameter
$b_\infty$ which is the distance between the pair of 5-branes, and
which is classically related to the ratio between the gauge
couplings. In the zero slope limit, this system becomes a $3+1$
dimensional Yang-Mills theory on ${\bf R}^{1,2} \times S^1$, with a
pair of domain wall defects localized on the $S^1$. We will discuss
here the further limit where we decouple the $3+1$ dimensional
dynamics and flow to a $2+1$ dimensional theory, by taking the radius
of the circle in the type IIB description to be small.

As in the previous section, one can follow the brane creation effects
by sliding $b_\infty$ past integer values. Each time the NS5-brane
crosses the $(1,k)$ 5-brane, $k$ D3-branes are created
\cite{Hanany:1996ie} (see figure \ref{figi}). Thus, for each slide by
a full cycle, the number of integer and fractional branes appears to
shift according to
\be
N_2  \rightarrow  N_2 + N_4, \qquad
N_4  \rightarrow  N_4 + k.
\label{nthreecascade}
\ee
In this setup, it is not necessary to shift $b$ by 2 to return to the
same configuration, since there are no fundamental matter fields, and
so there is no issue of keeping track of the gauge group with respect
to which they are charged.

One subtle issue here is that the brane configurations that we obtain
by these slides of $b_\infty$ at first sight seem to violate the
$s$-rule, which says that there can be at most $k$ D3-branes stretched
between an NS5-brane and a $(1,k)$ 5-brane
\cite{Hanany:1996ie,Kitao:1998mf,Bergman:1999na}. The original
configuration on the left of figure \ref{figi} is compatible with the
$s$-rule if $0 \leq N_4 \leq k$, but the other configurations seem to
have more than $k$ ``fractional D3-branes'' stretched between the
NS5-brane and the $(1,k)$ 5-brane. However, when the branes live on a
circle, we have to be careful about applying the $s$-rule.  In the
case that we have a fractional brane together with a regular brane, we
could interpret this either as one D3-brane stretched directly along
the segment from the NS5-brane to the $(1,k)$ 5-brane, and another
D3-brane wrapping the circle, or as a single D3-brane that winds
around the circle more than once. In other words, in the covering
space of the circle, we can have D3-branes that stretch between the
NS5-brane and different images of the $(1,k)$ 5-brane
\cite{Dasgupta:1999wx}, and the ``modified $s$-rule'' just tells us
that there can be at most $k$ D3-branes stretched between the
NS5-brane and a specific image of the $(1,k)$ 5-brane. Such a modified
$s$-rule is necessary for the low-energy physics to be independent of
$x^6$ positions of branes, even in highly supersymmetric
configurations such as those of the previous section. It also follows
from the derivation \cite{Bachas:1997sc,Bachas:1997kn} of the $s$-rule
(in a U-dual frame) by considering strings between two types of
D-branes (say, a D0-brane and a D8-brane), showing that they have a
single fermionic ground state, and noting that in supersymmetric
configurations this ground state is the only open string that can be
excited, and that this single fermionic ground state can be excited
(in a supersymmetric configuration) either zero times or once (and the
brane creation effect goes from one of these states to the
other). When we have a circle, we have different open string fermionic
ground states, that stretch between the two branes while also going
$n$ times around the circle, and each of these open string modes can
be excited either zero times or one time; each shift of $b_\infty \to
b_\infty+1$ excites one of these modes, but the modes excited by
different shifts have different winding numbers, so they can be
independently excited, and there is no violation of the $s$-rule.  The
fact that all the brane configurations of figure \ref{figi} seem to
have supersymmetric vacua has interesting implications, which we will
discuss in the next section.

In any case, the pattern \eqref{nthreecascade} of changes in $N_2$ and
$N_4$ implies that the Maxwell charges for this configuration must
take the form
\beq Q_2^{Maxwell} &=& N_2+ \left(N_4 - {k \over 2}\right) b_\infty + {k \over 2}
b_\infty^2,  \label{q2hw}\\
Q_4^{Maxwell} & = & \left(N_4 - {k \over 2}\right) + k b_\infty, \label{q4hw}
\eeq
(up to a possible overall shift of $Q_2$ by a multiple of $k$ which is
independent of $b_\infty$), as can be seen by substituting $b_\infty =
c+1$ and expanding according to the powers of $c$:
\beq Q_2^{Maxwell} & = & \left(N_2 + N_4\right) +
\left((N_4 +k) - {k \over 2}\right) c + {k \over 2} c^2,  \\
Q_4^{Maxwell} & = & \left(( N_4 +k )- {k \over 2}\right) + k c.
\eeq
Note that (up to terms which depend only on $k$ to which our analysis
above is not sensitive), this is equivalent to (\ref{maxwell1}) and
(\ref{maxwell2}) upon substituting
\be N_6 = 2k, \qquad N_{6+} = N_{6-} = k \ . \ee
A discussion similar to our discussion around equations
\eqref{qfourmax}, \eqref{qtwomax}, then implies that the Page charges
for this system are shifted as (up to a possible shift of $Q_2$ by a
multiple of $k$)
\beq Q_2^{Page} &=& N_2, \label{q2pagehw} \\
Q_4^{Page} & = & N_4 - {k \over 2} \label{q4pagehw} \ .
\eeq
In the rest of this section, we will provide evidence supporting this
expectation from the perspective of charge and flux quantization in
supergravity.

\subsection{Charge and flux quantization in supergravity duals of
Yang-Mills-Chern-Simons-Matter theories\label{s32}}

In this section, we examine the issues of charges and topology in the
dual gravitational description of the Yang-Mills-Chern-Simons-Matter
theory which arises as the zero slope limit of the brane
configurations described in the previous subsection.

The starting point is the metric of Lee, Weinberg, and Yi
\cite{Lee:1996kz} which describes the M-theory lift of overlapping
5-branes \cite{Gauntlett:1997pk}, namely, the configuration of the
previous subsection without the D3-branes. In the case of two
overlapping 5-branes with D5-brane and NS5-brane charges $(p_1,p_2)$
and $(\tilde{p}_1,\tilde{p}_2)$, it is given by
\be ds_{LWY}^2 = V_{ij} d\vec x_i d\vec x_j + (V^{-1})^{ij} R_i R_j (d \varphi_i+ A_i) (d
\varphi_j + A_j), \ee
where
\be V_{ij} = \delta_{ij} + {1 \over 2} {R_i p_i R_j p_j \over |R_1 p_1 \vec x_1+ R_2 p_2
\vec x_2| }
+
{1 \over 2} {R_i \tilde p_i R_j \tilde p_j \over |R_1 \tilde p_1 \vec x_1+ R_2 \tilde p_2
\vec x_2| } \ ,
\ee
$(i,j=1,2)$, $\vec{x}_1, \vec{x}_2$ are two 3-vectors, and $\varphi_i
\equiv \varphi_i + 2\pi$.  $A_1$ and $A_2$ are two one-forms whose
form we will not need explicitly.  For the theories we are interested
in, we have as in the previous subsection
\be (p_1,p_2) = (1,0), \qquad (\tilde p_1, \tilde p_2) = (1,k) \ . \ee
$R_1$ and $R_2$ are radii of $S^1 \times S^1$, which we identify as
the 6 and 11 directions, respectively. Therefore, they are related to
the field theory parameters by
\be R_1 = {2 \pi \alpha' \over L}, \qquad R_2 = R_{11} = g_s l_s = g_{YM2}^2 \alpha',
\label{scaling}\ee
where $L$ is the radius of the type IIB circle that the D3-branes live
on.  In order to decouple the excited string states and the $3+1$
dimensional dynamical degrees of freedom, and flow to a
Yang-Mills-Chern-Simons-Matter system in $2+1$ dimensions, we must
scale $\alpha', L\rightarrow 0$, which means taking $R_2\to 0$,
$R_2/R_1 \to 0$.

In the region near the core, this LWY geometry asymptotes to a simple
${\bf R}^8/{\bf Z}_k$ geometry \cite{Aharony:2008ug}.  This is the
regime where the geometry captures the dynamics of the ${\cal N}=6,8$
superconformal fixed point that the $2+1$ dimensional gauge theory
flows to in the IR.

Now we can consider repeating the analysis of the previous section,
replacing $ALE \times TN_k$ with the LWY space.  The strategy is to
add $N_2$ M2-branes and $N_4$ M5-branes wrapped on the torsion 3-cycle
of ${\bf R}^8/{\bf Z}_k$, and to explicitly construct the
(anti-)self-dual 4-form on the LWY geometry with the appropriate
boundary condition, as well as to solve for the harmonic function in
order to determine the warp factor. Unfortunately, despite the fact
that LWY, being a hyper K\"ahler manifold, has quite a bit of
structure, it is a far more complicated space than the $ALE \times
TN_k$ geometry (see e.g.  \cite{Hashimoto:2008iv} for the discussion
of the harmonic function on this space). Fortunately, for the purpose
of understanding the charges and their quantization, it is sufficient
to understand the broad topological structure of these spaces. We
leave the interesting mathematical exercise of finding the
(anti-)self-dual 4-form on LWY and the precise solutions dual to the
Yang-Mills-Chern-Simons theories for future work.

In order to better understand the topology of the LWY space, let us
first focus on the region near the origin. There, we find ${\bf R}^8
/{\bf Z}_k$ which is a cone whose base is $S^7/{\bf Z}_k$. We will
assume in the rest of this section that the geometry in the IR when we
add the additional branes is the $AdS_4\times S^7/{\bf Z}_k$ geometry
of \cite{Aharony:2008ug,Aharony:2008gk}; this is expected to be true
at least for $0 \leq N_4 \leq k$, and in the next section we will
suggest that it may have a wider range of validity. Reducing $S^7$ on
its Hopf fiber gives rise to a ${\bf CP}^3$ manifold, whose Betti
numbers are $(1,0,1,0,1,0,1)$. Because ${\bf CP}^3$ is also a K\"ahler
manifold, there is a K\"ahler form $J$, and its powers are dual to the
homology cycles on ${\bf CP}^3$, namely
\beq
\nonumber J &\leftrightarrow& {\bf CP}^1,  \\
\nonumber J \wedge J &\leftrightarrow& {\bf CP}^2, \\
J \wedge J \wedge J &\leftrightarrow& {\bf CP}^3.
\eeq
These forms are normalized such that
\be \int_{{\bf CP}^1} J = 1, \qquad  \int_{{\bf CP}^2} J \wedge J = 1, \qquad
 \int_{{\bf CP}^3} J \wedge J \wedge J = 1 \ . \ee
The RR 2-form field strength which one finds upon reducing M-theory on
LWY to type IIA string theory along the Hopf fiber is proportional to
$J$ :
\be F_2 = 2 \pi k R_{11} J. \ee

In the type IIA IR geometry, there is a ${\bf CP}^3$ at each radial
position, giving a foliation of the geometry transverse to the
D2-branes.  When this geometry is embedded inside the LWY geometry,
one expects an analogous foliation to continue into the UV, with a
corresponding two-form $J$.  Let us therefore assume that the
topological structure of ${\bf CP}^3$, its homology cycles and their
relation to the $J$'s, persist throughout the geometry.

In the type IIA language, the only 2-cycle on which one could have an
NS-NS 2-form potential associated with the distance $b_{\infty}$ is
the ${\bf CP}^1$. This suggests that the NS-NS 2-form potential should
interpolate between
\be B = (2 \pi)^2 \alpha'  b_\infty J \label{bUV}\ee
in the UV region of the geometry, and
\be B = (2 \pi)^2 \alpha'  b_0 J \ee
for some discrete $b_0$ (related to $N_4$ \cite{Aharony:2008gk}) in
the IR.

We now have most of the ingredients necessary to explore the issues of
charge quantization. Let us first consider the charge of D4-branes
wrapped on ${\bf CP}^1$ (which are the type IIA image of the
``fractional branes'' discussed above). The Page charge of these
branes, as in the previous section, is given by the flux of $\hat F_4$
through the dual cycle ${\bf CP}^2$. One important feature of the Page
charge is the fact that Gauss' law holds. So, computing this flux
should give the same result in the IR and in the UV. To compute the
Page charge in the IR, one evaluates
\be Q_4^{Page}={1 \over (2 \pi l_s)^3 g_s} \int_{{\bf CP}^2} (-\hat F_4) = {1 \over (2 \pi
l_s)^3 g_s} \int_{{\bf CP}^2} (-\tilde F_4 - B_2 \wedge F_2). \ee
In the IR, however, we do not expect $\tilde F_4$ to contribute (since
it vanishes in the $AdS_4\times S^7/{\bf Z}_k$ background), so we have
\be Q_4^{Page}= -{1 \over (2 \pi  l_s)^3 g_s} \int_{{\bf CP}^2}   B_2 \wedge F_2 =- k b_0 \
. \ee
Thus, we learn that \cite{Aharony:2008gk}
\be b_0 = -{Q_4^{Page} \over k} \label{quant}\ee
is quantized in units of $1/k$, since the Page charge $Q_4^{Page}$ is
expected to be quantized.

Once we know the Page charge, it is straightforward to compute the
Maxwell charges in the UV :
\beq
Q_2^{Maxwell} & = & Q_2^{Page} + b_\infty Q_4^{Page}+ {k \over 2}
b_\infty^2,\label{qmaxuv}\\
Q_4^{Maxwell} &=& Q_4^{Page} + k b_\infty,
\eeq
using (\ref{q2max}), (\ref{q4max}), and (\ref{bUV}). Note that these
results are in complete agreement with (\ref{q2hw}) and (\ref{q4hw}),
provided we identify the Page charges according to (\ref{q2pagehw})
and (\ref{q4pagehw}), in which $Q_4^{Page}$ is shifted with respect to
the naive integer quantized value by $k/2$. Let us now explain this
shift.

\subsection{Freed-Witten anomaly and a half integral shift in torsion flux}
\label{fwanom}

In sections \ref{s31} and \ref{s32}, we saw that the quantized charges
in the dual gravitational description of the
Yang-Mills-Chern-Simons-Matter system were compatible with the
expectation based on brane creation effects, provided that the charges
(\ref{q2pagehw}) and (\ref{q4pagehw}) were quantized with a
$k$-dependent fractional shift. Since the shift is in the Page charge
which satisfies Gauss' law, it suffices to understand this shift in
the IR (ABJM) limit.

In section \ref{s262} we also encountered a half-integral shift in the
D4-brane charge, which was due to the possibility of turning on an
integral self-dual flux on the worldvolume of D6-branes, leading to a
half-integral D4-brane charge.  Here, there are no D6-branes in
isolation, so the same argument can not be applied.

We can still try to use arguments similar to those of the previous
section by isolating a D6-brane. To do this, consider a probe D6-brane
wrapped on the ${\bf CP}^2$ cycle and extended along the field theory
directions.  Such a brane cannot be static, but this is not relevant
for our considerations; it will behave like a domain wall in $AdS_4$,
separating a region with $(k-1)$ units of D6-brane charge ($F_2$ flux)
and a region with $k$ units.

Because the manifold ${\bf CP}^2$ is not spin, the D-brane wrapped
on it is subject to the Freed-Witten anomaly \cite{Freed:1999vc}.
This anomaly arises because the path integral measure for the
fermions on such a D-brane is not globally well-defined by itself.
However, the total path integral measure is well-defined if the
worldvolume gauge field carries half-integral flux through the ${\bf
CP}^1$ homology 2-cycle of the ${\bf CP}^2$. This flux induces (if
there is no $B_2$ field) a $1/2$-unit of D4-brane Page charge and a
$1/8$-unit of D2-brane Page charge on the D6-brane, due to the
Chern-Simons coupling $C \wedge e^{B+F}$ on its worldvolume.  This
implies that when we go from the background with $k$ units of $F_2$ flux to the
background with $(k-1)$ units of $F_2$ flux by crossing the domain wall, the D4-brane
Page charge shifts from an integer to a half-integer or vice
versa. Since for $k=0$ we expect the charge to be an integer (in the
brane configuration; of course in this case there is no $AdS_4\times
S^7/{\bf Z}_k$ region), this implies that the fractional part of
$Q_4^{Page}$ should be equal to that of $k/2$, which is consistent
with equation (\ref{q4pagehw}).

While this result is in gratifying agreement with the analysis based
on brane configurations, it has one important consequence for the
dualities of ABJ and ABJM. This is because charge quantization now
imposes a condition (\ref{quant}) on $b_0$ which includes a shift by
1/2:
\be b_0 = - {N_4 \over k} + {1 \over 2}. \ee
We claim that the brane creation effects, as well as the careful
consideration of charge quantizations, suggest this shift in the value
of $b_0$ by one half, compared to the naive value assumed in
\cite{Aharony:2008ug,Aharony:2008gk}.\footnote{A similar shift by a
half can also be seen from the M theory perspective
\cite{Witten:1996md}, but we will not discuss this here.}  Note in
particular that even for the $U(N)\times U(N)$ theory of
\cite{Aharony:2008ug}, this implies that the $B_2$ field is
non-vanishing.

The same shift occurs also in the full solutions describing ${\cal
N}=3$ YMCS theories flowing to the ${\cal N}=6$ SCFTs. One class of
solutions of this type was constructed in \cite{Hashimoto:2008iv}. In
that paper it was assumed that $b=0$, but it is clear that the same
solutions exist with an additional $B_2$-field of the form
$b=-N_4/k+1/2$ for $N_4=0,\ldots,k-1$. Since the $B_2$-field is
constant, these solutions have $b_0=b_{\infty}$; from the point of
view of the discussion in this section, this means that they have a
specific relation between the value of $N_4$ and the ratio between the
two gauge couplings in the UV. In particular, for the case of $b=1/2$
these solutions describe the $U(N) \times U(N)$ theories with equal
gauge couplings in the UV, which gives a parity-invariant
theory. Generically, of course, there will be no such relation between
the UV gauge couplings and $N_4$; it would be interesting to construct
solutions for this more general case.\footnote{If we vary $b_\infty$
keeping $N_4$ fixed, there will be additional gravitational back
reaction which we discuss briefly in section \ref{cascadesec}.}

\subsection{Some consistency tests}

In \cite{Aharony:2008ug,Aharony:2008gk} several tests of the
conjectured correspondence between ${\cal N}=6$ Chern-Simons-matter
theories and type IIA string theory on $AdS_4 \times {\bf CP}^3$ {\it
without} the shift in $b_0$ by $1/2$ were presented. In this section,
we will show how each of these arguments remains valid also after the
shift of $b_0$ by $1/2$. Some of these tests can be regarded as
independent arguments for why $b_0$ should be shifted by $1/2$ (which
use the Freed-Witten anomaly, but do not directly use the shifted Page
charge).

\subsubsection{Domain wall brane charge}

In section \ref{fwanom}, the D6-brane wrapped on ${\bf CP}^2$ played
an important role. Let us consider the D2-brane ``brane charge''
associated with this object.

First, we need to account for the $1/8$ units of D2-brane charge which
were induced on the D6-brane, contributing to the Page charge. This
modifies (\ref{q2hw}) in the ABJM limit to read
\be Q_2^{Maxwell}= \left(N_2 + {k \over 8}\right) + b_0 \left(N_4 - {k \over 2}\right) + {k
\over 2} b_0^2  + Q_2^{anomaly} \ee
where\footnote{For $k=2$ and $N_4=0,1$, this evaluates to
$Q_2^{Maxwell}=N_2-1/16$ and $Q_2^{Maxwell}=N_2 + 3/16$, in agreement
with the computation of \cite{Sethi:1998zk}.}
\be b_0 = -{N_4 \over k} + {1 \over 2} \ . \ee
We have also introduced the contribution due to higher curvature terms
\be Q_2^{anomaly} = -{1 \over 24}\left( k - {1 \over k}\right)\ee
which was computed in \cite{Bergman:2009zh}.  These corrections,
depending only on $k$, do not affect the brane manipulation argument
from the previous section.  Nonetheless, they turn out to be critical
in some of the charge quantization issues as was the case in
\cite{Freed:1999vc}.

In particular, the Page charge, which is related to the Maxwell charge
(\ref{q2max}) as
\beq Q_2^{Page} &=& {1 \over (2 \pi l_s)^5 g_s}\int_{{\bf CP^3}}   \hat F_6 \cr
&  = &
{1 \over (2 \pi l_s)^5 g_s} \int_{{\bf CP^3}} \left[ * \tilde F_4 - B_2 \wedge (-\hat F_4) - {1
\over 2} B_2 \wedge B_2 \wedge F_2 \right], \label{q2max2}
\eeq
is now shifted to take on a non-integer value (at leading order in
large $k$)
\be Q_2^{Page} = N_2 + {k \over 8} - {k \over 24} = N_2 + {k \over 12} \ . \ee

The D2-brane brane charge on the domain wall of section \ref{fwanom},
which shifts the value of $k$ by one, can readily be computed:
\be Q_2^{Brane} = Q_2^{Maxwell}(N_2, N_4, k)  -  Q_2^{Maxwell}(N_2, N_4, k-1)  =  {1 \over
2} \left({N_4 \over  k}\right)^2 - {1 \over 24} + {\cal O}(k^{-1})  \ \label{diffmax},
\ee
where we drop terms subleading in $1/k$ for the purpose of comparing
this result with the probe approximation. This is indeed in agreement
with the expected contributions from the terms in the worldvolume
action \cite{Green:1996dd}
\be S^{R}_6 \sim \int C \wedge e^{F+B} \wedge \sqrt{
{{\hat {\cal A}}(4 \pi^2 R_T)  \over {\hat {\cal A}}(4 \pi^2 R_N) }}\ . \ee
Here,
\be \hat {\cal A} = 1 - {1 \over 24} p_1  + {1 \over 5760} (7 p_1^2 - 4 p_2)\ee
is the ``A-roof'' genus, which is expressed in terms of Pontryagin
classes
\be p_1 =-{1 \over 2 (2 \pi)^2} \tr R^2, \qquad p_2 = {1 \over 8 (2 \pi)^4} [(\tr R^2)^2 - 2
\tr R^4]\ . \ee

It is useful to note that for ${\bf CP^2}$ embedded into ${\bf CP^3}$,
\beq  \int_{{\bf CP^2}} p_1(T) &=& 3, \\
\int_{{\bf CP^2}}  p_1(N) &=& 1, \eeq
are the first Pontryagin classes for the tangent and the normal
bundles, respectively \cite{Eguchi:1980jx}.

The resulting D2-brane charge is
\be Q_2^{Brane} = Q_2^{WZ} + Q_2^{R},\ee
where (recalling that the D6-brane has a half-integer flux of its
gauge field $F$)
\be Q_2^{WZ}={1 \over (2 \pi l_s)^4 } \int_{{\bf CP^2}} \left[ {1 \over 2}  (B_2+F) \wedge (B_2+F) \right]
= {1 \over 2} \left({N_4 \over  k}\right)^2,\ee
and the higher curvature terms
\be  Q_2^R= \int_{{\bf CP^2}} {1 \over 48}
(- p_1(T)+p_1(N))
 = -{1 \over 24}  \ ,  \ee
giving rise to a result matching (\ref{diffmax}).\footnote{AH thanks
Yuji Tachikawa for a conversation related to this issue.}

\subsubsection{Di-baryon vertex}

A D4-brane wrapping the ${\bf CP}^2$ cycle in $AdS_4 \times {\bf
CP}^3$ describes a particle-like object in $AdS_4$. The Ramond-Ramond
fluxes induce an electric field on the D4-brane worldvolume, which
must be canceled by some number of fundamental strings ending on the
D4-brane, as for the baryon vertex in $AdS_5 \times S^5$
\cite{Witten:1998xy}. The number of such strings follows from the CS
couplings on the D4-brane worldvolume, which are proportional to
\beq S^{WZ} & \sim &  -\int_{{\bf R} \times {\bf CP}^2} A \wedge (F_4 + (B_2+F) \wedge
F_2)\cr
&=&-\int_{{\bf R} \times {\bf CP}^2} A \wedge (\hat F_4 + F \wedge F_2).\eeq
This object was identified in \cite{Aharony:2008ug,Aharony:2008gk}
with a di-baryon, made by contracting $N_2$ bi-fundamentals to a
singlet using the epsilon symbol in $SU(N_2)$; in the $U(N_2)\times
U(N_2)$ theory such an object can be a singlet also of the other
$SU(N_2)$ group (so that it only carries some overall $U(1)_B$
charge), while in the $U(N_2)\times U(N_2+N_4)$ theory such an object
naturally lives in the $N_4$'th anti-symmetric product of fundamental
representations of $SU(N_2+N_4)$. Thus, we expect to need to have
$N_4$ strings ending on this object.

Therefore, we expect
\be N_4 = {1 \over (2 \pi l_s)^3 g_s} \int_{\bf{CP}^2} (-\hat F_4 - F \wedge F_2)  \ . \ee
Recall that
\be {1 \over (2 \pi l_s)^3 g_s} \int (-\hat F_4) = Q_4^{Page} =  - k b_0. \ee
Moreover, following the Freed-Witten argument presented in the
previous section, we know that the brane wrapped on ${\bf CP}^2$
should have a $half$-integer gauge field flux on its worldvolume.  The
minimal possibility for the gauge flux is\footnote{Other possible
values of the flux correspond to bound states of these D4-branes with
D2-branes wrapped on ${\bf CP}^1$, which are vertices which have $k$
strings ending on them \cite{Aharony:2008ug}.} is
\be {1 \over (2 \pi l_s)^3 g_s} \int  (-F \wedge F_2) = {k \over 2}, \ee
which suggests that $b_0$ must be
\be b_0 = -{N_4 \over k} + {1 \over 2}.\ee
Thus, the di-baryon vertex is consistent with our arguments above.

\subsubsection{Baryon vertex}

The D6-brane wrapped on ${\bf CP}^3$ was interpreted in
\cite{Aharony:2008ug} as a baryon vertex, on which $N_2$ strings
corresponding to external sources in the fundamental representation of
$U(N_2)$ can end. (Of course, we also expect to have an analogous
object with $N_2+N_4$ strings ending on it, but this is simply the
bound state of the D6-brane with the D4-brane described above, or,
equivalently, a D6-brane with one unit of gauge field flux on ${\bf
CP}^1$.) Since ${\bf CP}^3$ is a spin manifold, such a D6-brane does
not experience the effects of the Freed-Witten anomaly, so the minimal
energy configuration has a vanishing worldvolume flux.  Denoting the
worldvolume gauge potential by $A$, and its field strength by $F$, the
terms in the worldvolume action responsible for determining the number
of fundamental strings ending on the D6-brane are
\beq S_{D6}^{WZ} & \sim & - \int \left(A_7 + A_5 \wedge (F + B_2) + {1 \over 2}  A_3 \wedge
(F+B_2) \wedge (F+B_2) \right.\cr
&& \left.+ {1 \over 3!} A_1 \wedge (F + B_2) \wedge (F + B_2) \wedge (F + B_2)  \right)\cr
&=& \int \left[ - A_7 +  A \wedge \left(F_6 + A_3 \wedge H_3 + B_2 \wedge  F_4 + {1 \over 2}
B_2 \wedge B_2 \wedge F_2 \right) \right] \cr
& = & \int (-A_7 + A \wedge \hat F_6 ). \eeq
where we dropped the terms proportional to $H_3$, as they vanish. We
recognize the D2-brane Page charge
\be Q_2^{Page} = N_2 + {k \over 12} \label{baryon-anom}\ee
as  giving rise to an anomalous worldvolume electric charge.

The contribution from the higher curvature terms, on the other hand,
is given by
\beq  S^R_{D6} & \sim &  - \int A_1 \wedge (F+B) \wedge {1 \over 48}  (p_1(N) - p_1(T))\cr
& = &   \int A \wedge F_2  \wedge {1 \over 48}  (- p_1(T)+p_1(N)). \eeq
We can evaluate this using
\be F_2 = 2 \pi k g_s l_s J \ , \ee
and
\beq   \int_{{\bf CP}^3} J  \wedge   p_1(T) & = &  4, \\
 \int_{{\bf CP}^3} J  \wedge   p_1(N) & = & 0,
\eeq
which are the periods over ${\bf CP^3}$ of the first Pontryagin
classes of the tangent and normal bundles, respectively, wedged with
the K\"ahler form $J$ \cite{Eguchi:1980jx}. We obtain
\be {1 \over 2 \pi g_s l_s}  \int_{{\bf CP}^3} F_2  \wedge {1 \over 48}  (- p_1(T)+p_1(N)) =
-{k  \over 12} . \ee
This cancels the non-integer part of (\ref{baryon-anom}), giving rise
to the net charge $N_2$, which is precisely the expected number (and,
in particular, it is an integer).  This test is therefore unaffected
by the shift in $b_0$ by 1/2.

\subsubsection{Parity}
\label{parity_sec}

An important consistency check of our proposal is to show that the
field theory and its gravity dual transform under parity in compatible
ways.

In the absence of fractional branes, the ABJM superconformal field
theory is invariant under the standard parity action of 3-dimensional
gauge theory (reflecting one of the spatial coordinates, say $x^2$).
Parity changes the sign of the two Chern-Simons terms, but because the
gauge group is $U(N_2)_k \times U(N_2)_{-k}$ this maps the theory to
itself.  On the gravity side, M theory and type IIA string theory both
admit a parity action, but this acts also as a change of sign for some
of the background fields.  In particular, in type IIA, parity takes
$B_2 \rightarrow -B_2$. Taking $B_2 =(2 \pi)^2 \alpha' b_0 J$ on the
${\bf CP}^1$ cycle in ${\bf CP}^3$ as above, an obvious
parity-invariant choice is $b_0=0$.  There is however an alternative
choice, $b_0=1/2$. Under parity, this maps to $b_0 \rightarrow -1/2$,
but since type IIA string theory is also invariant under $b_0 \sim b_0
+ 1$, the choice $b_0=1/2$ is actually also invariant.

When fractional branes are included, the unequal ranks of the gauge
groups imply that parity is not a symmetry of the theory, but rather
it takes the theory into a different ``parity-dual'' theory, as
pointed out in \cite{Aharony:2008gk}. The inverse of the ``slide
transformation'' \eqref{nthreecascade} takes the $U(N_2)_k\times
U(N_2+N_4)_{-k}$ theory to a $U(N_2+k-N_4)_k\times U(N_2)_{-k}$
theory, which was argued in \cite{Aharony:2008gk} to be a dual
description of the same SCFT, and the naive parity transformation then
takes this to a $U(N_2)_k\times U(N_2+k-N_4)_{-k}$ theory.  So, the
parity transformation was conjectured to exchange $N_4\to k-N_4$
without changing $k$.

In the gravity dual, it was argued in \cite{Aharony:2008gk} that each
fractional brane shifts $b_0$ by $1/k$, and it was assumed there that
$b_0$ vanishes for $N_4=0$, leading to $b_0=N_4/k$. Under parity
(again acting as $x^2 \rightarrow -x^2$) this maps to $b_0=-N_4/k$
which is equivalent by a large gauge transformation to
$b_0=(k-N_4)/k$, consistent with the field theory arguments above.

With the shifted quantization condition $b_0 = N_4/k -1/2$, we see
that the parity transformation $b_0 \rightarrow -b_0$ is again
equivalent to sending $N_4 \rightarrow k-N_4$, this time without
needing to perform any large gauge transformation shifting $b_0$.
This shows that our proposal still admits a natural (and possibly
simpler) action of parity.

\section{Duality cascades in ${\cal N}=3$ Yang-Mills-Chern-Simons theories\label{cascadesec}}

The relation described above between the brane configurations of
figure \ref{figi} suggests that perhaps different $U(N_2)\times
U(N_2+N_4)$ ${\cal N}=3$ supersymmetric Yang-Mills-Chern-Simons
theories could be equivalent at low energies. The simplest expectation
may be that all of these theories flow to $U(N_2)\times U(N_2+N_4)$
${\cal N}=6$-superconformal Chern-Simons-matter theories, and that the
superconformal theories related by \eqref{nthreecascade} are all
equivalent. However, it was argued in \cite{Aharony:2008gk} that such
superconformal theories exist only when $|N_4| \leq k$, since in other
cases one can obtain on their moduli space (by moving out the $N_2$
``regular branes'') pure ${\cal N}=3$ supersymmetric $U(N_4)_k$
theories with $|N_4| > k$, that are believed
\cite{Kitao:1998mf,Bergman:1999na,Ohta:1999iv} to break
supersymmetry. Thus, at most one cascade step of \eqref{nthreecascade}
can really relate ${\cal N}=6$ superconformal theories, which is
precisely the one described in section \ref{parity_sec} above.

Nevertheless, we can still conjecture that when we have a
$U(N_2+N_4)_k\times U(N_2+2N_4+k)_{-k}$ Yang-Mills-Chern-Simons theory
with $0 < N_4 < k$, related by a single cascade step
\eqref{nthreecascade} to a $U(N_2)_k\times U(N_2+N_4)_{-k}$ theory
which is believed to flow to a ${\cal N}=6$ superconformal fixed
point, then this theory flows in the IR to the same fixed point. The
original theory classically has an $8(N_2+N_4)$-dimensional moduli
space, but, as mentioned above, assuming that this moduli space
persists in the quantum theory leads to a contradiction. The
conjecture above implies that the quantum moduli space is smaller, and
is only $8N_2$-dimensional. This has a nice interpretation using the
modified $s$-rule discussed in the previous section. This $s$-rule
tells us that the naive brane connections of the $U(N_2+N_4)\times
U(N_2+2N_4+k)$ theory, which involve $(N_2+N_4)$ regular branes going
around the circle, do not lead to a supersymmetric
configuration. However, there is another way to connect the branes, in
which there are only $N_2$ regular branes (and the other $N_4$ regular
branes join together with a fractional brane to give a fractional
brane of higher winding number), and this connection does not violate
the $s$-rule (and, thus, is expected to lead to supersymmetric
vacua). Similar arguments suggest that the $U(N_2+n N_4+n(n-1)k
/2)_k\times U(N_2+(n+1) N_4 + n(n+1) k / 2)_{-k}$ theory may also flow
to the same ${\cal N}=6$ superconformal fixed point as the
$U(N_2)_k\times U(N_2+N_4)_{-k}$ theory, for any integer $n$ (when
$|N_4| \leq k$), and that, correspondingly, it has a smaller quantum
moduli space (which agrees with the modified $s$-rule). This flow
involves a sequence of dualities reducing $N_2$ and $N_4$, which are
similar to Seiberg dualities, so the final picture would be very
similar to the cascade of Klebanov and Strassler, except that here the
cascade can smoothly end with a finite rank at high energies, and its
IR end is given by a superconformal theory rather than a theory with a
mass gap.\footnote{The picture of different ways to connect brane
configurations on a circle, leading to different dimensional moduli
spaces, may also be applied to the Klebanov-Strassler cascade
\cite{Klebanov:2000hb,Strassler:2005qs,Dymarsky:2005xt}. The naive
brane connection of the $SU(N)\times SU(N+M)$ theory in that case
leads to a $6N$-dimensional moduli space, but one could also connect
$M$ regular branes to $M$ fractional branes to form $M$ fractional
branes winding once around the circle, or connect $2M$ regular branes
to $M$ fractional branes to form $M$ fractional branes winding twice
around the circle, etc. This suggests that the same theory may also
have moduli spaces of dimensions $6(N-M)$, $6(N-2M)$, etc., and the
existence of such branches of the moduli space is indeed confirmed by
a careful analysis of the quantum moduli space of this theory
\cite{Dymarsky:2005xt}. Note that the match between these pictures is
not precise, since in principle one could also connect just one
regular brane to one fractional brane, and there is no corresponding
branch of the moduli space found in
\cite{Dymarsky:2005xt}. Nevertheless, this example suggests that in
some cases one can interpret different branches of the moduli space as
corresponding to different brane connections, and we would like to
claim that the same thing happens in our case.}

Note that this conjecture still does not imply that all
$U(N_2)_k\times U(N_2+N_4)_{-k}$ theories have supersymmetric vacua;
when we start from $N_4 > k$ we can always try to perform cascade
steps which are the inverse of \eqref{nthreecascade}, but when $N_4$
is too large we get to $N_2 \leq 0$ before we get to $N_4 < k$, and in
these cases there is no possible superconformal field theory to flow
to in the IR. Thus, we suggest that these cases (and, in particular,
the $U(N_4)_k$ theory arising for $N_2=0$ when $N_4>k$, as discussed
in \cite{Kitao:1998mf,Bergman:1999na,Ohta:1999iv}) dynamically break
supersymmetry\footnote{For the theories with $N_2>0$ another
possibility is that they have no vacuum, and have a runaway behavior
along their classical moduli space.}.  However, in some cases this
breaking may happen in the IR after the theory first undergoes a
series of duality cascades as described above.

Of course, the arguments for the duality cascade that we gave above
were circumstantial and involved various assumptions and
analogies. The cascade is consistent with all the usual consistency
checks for supersymmetric dualities, once we assume that the moduli
space of ${\cal N}=3$ supersymmetric Yang-Mills-Chern-Simons theories
can be modified in the quantum theory as described above. Since the
field theories in question are always strongly coupled, it seems that
the best way to test if such a cascade exists or not is to construct
the dual string theory solutions, analogous to the one of
\cite{Klebanov:2000hb}, assuming that they are weakly curved and can
be described by supergravity.  This issue is currently under
investigation \cite{nextgreatpaper}.

\section*{Acknowledgements}

AH would like to thank Sergey Cherkis for extended discussions on
related topics over many years, and N. Itzhaki for discussions about
the D1-D3 system. We would also like to thank Oren Bergman, Stefano
Cremonesi, Keshav Dasgupta, Paolo Di Vecchia, Davide Gaiotto, Daniel
Jafferis, Juan Maldacena, and Yuji Tachikawa for useful discussions
and comments on a draft of this manuscript. The work of AH and PO was
supported in part by the DOE grant DE-FG02-95ER40896.  The work of OA
was supported in part by the Israel-U.S. Binational Science
Foundation, by a center of excellence supported by the Israel Science
Foundation (grant number 1468/06), by a grant (DIP H52) of the German
Israel Project Cooperation, by the European network
MRTN-CT-2004-512194, and by Minerva.  The work of SH was supported by
FNU via grant number 272-06-0434.

\appendix

\section{Flux and potential quantization in the D1-D3 system\label{appA}}

In this appendix we review the status of flux and charge quantization
in the D1-D3 bound state system in type IIB string theory. Consider a
system consisting of $N_1$ D1-branes extended along the 01 directions
and $N_3$ D3-branes extended along the 0123 directions, with the 23
directions compactified on $T^2$ with some non-vanishing $B_2$-field
along this $T^2$. We will take the D1-branes to have melted into the
D3-branes, so that there is a non-vanishing $F_{23}$ gauge field flux
on the worldvolume of the D3-branes.  With a non-zero $B_{23}$, one
expects a shift in the D1-brane charge carried by the D3-branes, from
$F_{23}$ to $F_{23}+B_{23}$. The energy density of the branes is also
shifted due to the presence of $B_2$ in the $\sqrt{-\det(G+B+F)}$ DBI
action of the D3-branes, affecting their ADM mass.

To see the gravitational back-reaction of this, consider the type IIB
supergravity solution for such a D1-D3 bound state
\cite{Russo:1996if}:
\beq
ds^2_{str} &=& f^{-1/2} [ - dx_0^2 + dx_1^2 +{h}(dx_2^2 + dx_3^2 )]
+ f^{1/2} ( dr^2 + r^2 d \Omega^2_5 )\ ,
\cr
f &= & 1 + {\alpha'^2 R^4\over r^4 } ~, \qquad
h^{-1} =\sin^2\theta f^{-1}+\cos^2\theta \ ,
\\
B_{23} &=& \left(B_\infty-{\sin\theta\over \cos\theta } \right)+{\sin\theta\over \cos\theta
}\ \ f^{-1} h \ , \label{B}
\cr
e^{2\phi} &=& g_s^2 h \ ,\qquad
F_{01r} = {1 \over g_s} \sin\theta\ \partial_r f^{-1} \ , \cr
F_5 &=& {\tilde F}_5 + *{\tilde F}_5,
\qquad
{\tilde F}_{0123r} = {1 \over g_s} \cos\theta \ h \ \partial_r f^{-1} \ .
\eeq
The parameters $R$ and $\theta$ will be related to the charges below.
We included in the solution a possible non-vanishing $B_{23}$ at
infinity by hand, which we expect to correspond to the possibility of
turning on a $B_2$ field on the torus asymptotically far away from the
branes.  At the level of solving the SUGRA equations of motion, this
has no effect. On the other hand, one must explore how the charge
quantization conditions are affected.

Computation of the self-dual 5-form flux in this background gives
\be {1 \over (2 \pi l_s)^4} \int_{S^5} F_5 = {1 \over (2 \pi l_s)^4 g_s} 4 \alpha'^2 R^4
\cos \theta \Omega_5 =
N_3\ , \ee
where $\Omega_5$ is the volume of a unit 5-sphere, which implies
\be  R^4 \cos \theta  =  4 \pi g_s N_3\,.  \label{rel1a} \ee

With regards to the D1-brane charge, on the other hand, the modified
equation of motion
\be d * F_3 = H \wedge F_5 \ee
implies that $* F_3$ is not closed and will not respect Gauss'
law. However, the modification
\be \hat F_7 = * F_3 - B \wedge F_5 \label{D1page} \ee
will give rise to a closed non-gauge-invariant 7-form.  One can then
either compute the Maxwell charge
\be Q_1^{Maxwell} = {1 \over (2 \pi l_s)^5} \int_{T^2 \times S^5}  * F_3   =
{N_3 R_1^2 \over \alpha'} \sin \theta \label{maxwell}
\ee
or the Page charge
\be Q_1^{Page}={1 \over (2 \pi l_s)^5} \int_{T^2 \times S^5}  \hat F_7  = {N_3 R_1^2 \over
\alpha'} (\sin \theta - B_\infty \cos \theta) = N_1, \ee
where $R_1$ refers to the radius of the $x_2$ and $x_3$ coordinates
and is not to be confused with $R$, which is a parameter in $f$.

Combined with (\ref{rel1a}), this implies
\beq R^4 & = & 4 \pi g N_3 \sqrt{\frac{\left(B_\infty N_3
   R_1^2+\alpha' N_1\right)^2}{N_3^2
   R_1^4}+1}, \label{ans1}\\
\tan \theta &=& {\alpha'  \over R_1^2}{N_1 \over N_3} +B_\infty, \label{ans2}\eeq
which prescribes exactly how $R$ and $\theta$ are discretized due to
the discreteness of $N_1$ and $N_3$ which are integer-quantized.  The
Maxwell charge (\ref{maxwell}) with $R$ and $\theta$ expressed in
terms of $N_1$ and $N_3$ has the form
\be {64 \pi \alpha'^3 \over R_1^2} \left(N_1 + {B_\infty R_1^2  \over \alpha'}N_3 \right).
\ee
We see that the natural periodicity of $B_\infty$ in units of
$\alpha'/R_1^2$ appear. A useful dimensionless parameter to define,
therefore, is
\be \chi = {R_1^2\over \alpha'} B_{\infty}. \ee

Suppose we now consider taking the near-horizon limit keeping $\chi$,
\be U \equiv {r \over \alpha'} \ , \ee
and $R_1$ fixed.  Then, the geometry approaches
\be {ds^2\over \alpha'} = {U^2 \over R^2} (-dt^2 + dx^2+dy^2+dz^2) + {R^2 \over U^2} (dU^2 +
U^2 d \Omega_5^2), \ee
with
\be R = 4 \pi g_s N_3 \ee
and
\be B_{23} = - {\alpha' \over R_1^2} {N_1 \over N_3} \ . \ee
Note that $\chi$ has decoupled from $B_{23}$, which now takes on
values quantized in units of ${\alpha' / R_1^2 N_3}$.  This shows that
in the $AdS_5 \times S^5$ background (on a torus), $B_{23}$ is not
allowed to take arbitrary values at the quantum level. One can also
think of the degree of freedom contained in $\chi$ as a singleton,
which decouples in the limit where one focuses only on the $SU(N_3)$
part of the dynamics of the stack of D3-branes. Changes in $B_{23}$
are locked with changes in $N_1$ through the presence of Chern-Simons
terms, and the latter are only allowed to take on discrete
values. Note also that these issues are strictly quantum and are
invisible in the large $N_3$ limit.

Once we identify (\ref{D1page}) as the Page charge, it is
straightforward, via a chain of T-dualities, to obtain
(\ref{d4page}). The only difference between the D1-D3 system and the
D6-brane wrapped on $ALE$ is the orientation of the $B_2$-field.
Otherwise, the issues of induced charges and their quantization are
identical.

One interesting question remains. How does one think of the Page
charge (\ref{d4page}) when lifted to M theory? The RR flux $F_2$ lifts
in M-theory to a geometric flux. A natural candidate expression,
borrowing the notation of \cite{Kachru:2002sk,Shelton:2005cf}, is to
write
\be \hat G_4=G_4 + f \wedge C_3, \ee
where
\be f \wedge C = f^a_{bc} C_{a d e} dx^b \wedge dx^c \wedge dx^d \wedge dx^e, \ee
and to use $\hat G_4$ to define the Page flux of the fractional
M5-brane.  It would be interesting to verify if this expression is
correct.

\section{Conventions\label{appB}}

\subsection{Differential geometry conventions}

Much of the material presented in this article depends sensitively on
the details of sign conventions for Type II supergravity. In this
appendix, we summarize the conventions which we follow.

\bl The signature of space-time is taken to be $(-,+,\ldots,+)$.

\bl Differential forms are denoted $F_p = {1 \over p!} F_{i_1 \ldots i_p} dx^{i_1} \wedge
\ldots dx^{i_p}$.

\bl Hodge duals are defined by
\be *F_{p} = {1 \over (D-p)!} F_{i_1 \ldots i_p} \epsilon^{i_1 \ldots i_p}{}_{i_{p+1} \ldots
i_D} dx^{i_{p+1}} \wedge \cdots \wedge dx^{i_{D}}, \ee
where
\be \sqrt{-\det g} \, \epsilon^{01 \ldots (D-1)} = 1 \ . \ee

\bl $|F|^2 \equiv F_{\mu_1 \ldots \mu_p} F^{\mu_1 \ldots \mu_p} = -F \wedge * F$.

\bl As a result of the definition of the Hodge dual operator $*$,
\be * * F_p =-F_p \ee
for even $p$-form field strengths in type IIA supergravity.

\subsection{Supergravity conventions}

We consider the type IIA supergravity Lagrangian
\be S = {1 \over 2 \kappa_{10}^2} \left(S_{NSNS} + S_{RR} + S_{CS} + S_{DBI} + S_{WZ}
\right),\ee
where the bulk terms are
\beq S_{NSNS} &=& \int d^{10} x \sqrt{-g} e^{-2 \phi} \left(R + 4 \partial_\mu \phi
\partial^\mu \phi - {1 \over 2} \left| H_3 \right|^2\right),\\
S_{RR} &=& -{1 \over 2} \int d^{10} x \sqrt{-g}  \left(|F_2|^2 + |\tilde F_4|^2\right),  \\
S_{CS} & = & {1 \over 2} \int B_2 \wedge F_4 \wedge F_4, \label{Scs}
\eeq
and the action on D-branes is
\beq
S_{DBI} & = & - (2 \kappa_{10}^2 T_p) \int d^{p+1} \sqrt{-\det(g + B + F)}, \\
S_{WZ} & = & -  s \int A \wedge  e^{B_2}  \wedge * j_{p+1}^{Dp} \ , \label{Swz}
\eeq
with $s = +1$ for a brane and $s = -1$ for an anti-brane.
The constants and currents are normalized in terms of the string
coupling and the string scale as
\beq 4 \pi \kappa_{10}^2 &=& (2 \pi l_s)^8 g_s^2 \\
2 \pi \kappa_{10}^2 T_p &=& {(2 \pi l_s)^{7-p} g_s} \\
j_{p+1}^{Dp} &=& (2 \pi l_s)^{7-p} g_s\delta^{9-p}(\vec{y})\, dx^0 \wedge dx^1 \ldots \wedge dx^p \ , \eeq
where $\vec{y}$ denotes the transverse dimensions to the Dp-brane.
This set of conventions has several implicit implications which we
elaborate below.

\bl The type II supergravity D$p$-branes with positive charge will
have RR $p+1$-form potential
\be A_{p+1} = - H^{-1} dx^0 \wedge \cdots \wedge dx^{p+1}, \ee
where $H \sim r^{-(7-p)}$ and is positive for large $r$. An isolated
brane will satisfy the linear relation
\be d * F_{p+1} = s * j_{p+1}^{Dp}. \ee

\bl We take the gauge-invariant type IIA 4-form to be
\be \tilde F_4 = d A_3 + d B_2 \wedge A_1. \label{gi4form} \ee

\bl The form of $S_{WZ}$ implies that a D$(p+2)$-brane in a presence
of $B_2$-field with positive $\int B_2$ will induce a positive
D$p$-brane charge.

\bl The presence of D4-branes and D6-branes implies that form fields
$A_5$ and $A_7$ are implicitly part of the $S_{WZ}$ term. They are
understood as arising from generalizing the free part of $S_{RR}$ so
that
\be S_{RR} = -{1 \over 4 \kappa_{10}^2 } \int d^{10} x \sqrt{-g}  \left(|F_2|^2 + |F_4|^2 +
|F_6|^2 + |F_8|^2\right),
\ee
but with the constraint
\be F_2 = * F_8, \qquad F_4 = -* F_6. \label{dualrel} \ee
The constraint should not be viewed as being imposed directly on the
action. Instead, note that the equations of motion
\beq d * F_8 &=& s * j_7^{D6},  \\
 d * F_6 &=& s * j_5^{D4},
\eeq
can be interpreted as imposing the following condition for the fluxes
\beq \int_{S_2} F_2 &=& \int_{V_3} d F_2 =  \int_{V_3}  d * F_8 = s\int_{V_3}   *  j_7^{D6},
\\
\int_{S_4}(- F_4) &=& \int_{V_5} d (-F_4) =  \int_{V_5}  d * F_6 = s \int_{V_5}   *
j_5^{D4},
\eeq
with $S_i = \partial V_{i+1}$. The seemingly unnatural choice of signs
in (\ref{dualrel}) can be motivated by considering the extension of
the consideration above to type IIA supergravity, including the
non-linear effects. They are constrained by the choice of the sign of
the type IIA Chern-Simons term (\ref{Scs}), the form of the
gauge-invariant four-form (\ref{gi4form}), and the choice of the sign
of $B_2$ in $S_{WZ}$ (\ref{Swz}), as can be seen from the following
considerations.

The Maxwell equation and the Bianchi-identity for the 2-form and the
4-form field strengths take the form
\beq
d F_2 & = &  * j^{brane}_7  = s_6 * j^{D6}_7,  \label{rel1}\\
 - d \tilde F_4 &=&     * j^{brane}_5+ H_3 \wedge F_2
 =  s_4 * j^{D4}_5+ d(B_2 \wedge F_2), \label{rel2}\\
d * \tilde F_4 & = &  * j^{brane}_3 - H_3 \wedge \tilde F_4
=  s_2 * j^{D2}_3+d\left(B_2 \wedge (- \tilde F_4) - {1 \over 2} B_2 \wedge B_2 \wedge F_2\right), \label{rel3}\\
d * F_2 & = &   * j^{brane}_1+H_3 \wedge * \tilde{F}_4
=  s_0 * j^{D0}_1\cr
&& \qquad +d\left(B_2 \wedge (*\tilde{F}_4)
- {1 \over 2} B_2 \wedge B_2 \wedge (-\tilde F_4) + {1 \over 6} B_2 \wedge B_2 \wedge B_2 \wedge F_2
\right), \label{rel4}
\eeq
where $* j^{brane}_{p+1}=\sum_{q=p}^6\left(s_q e^{B_2}\wedge
*j^{Dq}_{q+1}\right)_{9-p}$ denotes the ``brane charge'' current, and
$s_q$ denotes the orientation of the D$q$-branes.  If the surface
integrals of the right-hand sides of the above expressions are to be
interpretable as the supergravity manifestation of induced charges due
to $S_{WZ}$, the only consistent identification is (\ref{dualrel}).

In this convention, the D-brane charges are computed as a surface
integral
\beq Q_6^{Page} &=& \int F_2, \\
Q_4^{Page} &=& \int (- \tilde F_4)- B_2 \wedge F_2, \\
Q_2^{Page} &=& \int * \tilde F_4 - B_2 \wedge (-\tilde F_4)
+ {1 \over 2} B_2 \wedge B_2 \wedge F_2, \\
Q_0^{Page} &=& \int *  F_2 - B_2 \wedge (* \tilde F_4)
+ {1 \over 2} B_2 \wedge B_2 \wedge (-\tilde F_4) - {1 \over 6} B_2 \wedge B_2 \wedge B_2 \wedge F_2
\ .
\eeq
The relations (\ref{rel2}) and (\ref{rel4}) are closely tied to the
form of the gauge-invariant four-form (\ref{gi4form}). The relation
(\ref{rel3}), on the other hand, is connected to the sign of the
Chern-Simons term (\ref{Scs}).  Had the sign of the Chern-Simons term
(\ref{Scs}) been switched, it would have been natural, instead, to use
opposite signs in \eqref{dualrel}.

\bl The gauge-invariant 4-form (\ref{gi4form}) arises from dimensional
reduction of the M-theory 4-form by the ansatz
\be C_3 = A_3 - B_2 \wedge dx^{11}, \ee
and lifts to the 11-dimensional action
\be S_{11} = {1 \over 2 \kappa_{11}^2} \int d^{11}x\,  \sqrt{-g} \left(R - {1 \over 2}
|G_4|^2 \right) - {1 \over  2 \kappa_{11}^2} \int {1 \over 6}   C_3 \wedge G_4 \wedge G_4 \
. \ee

\bibliography{frac}\bibliographystyle{utphys}

\providecommand{\href}[2]{#2}\begingroup\raggedright\begin{thebibliography}{10}

\bibitem{Bagger:2007jr}
J.~Bagger and N.~Lambert, ``{Gauge symmetry and supersymmetry of multiple
  M2-branes},'' {\em Phys. Rev.} {\bf D77} (2008) 065008,
\href{http://www.arXiv.org/abs/arXiv:0711.0955}{{\tt arXiv:0711.0955}}.

\bibitem{Gustavsson:2007vu}
A.~Gustavsson, ``{Algebraic structures on parallel M2-branes},''
\href{http://www.arXiv.org/abs/arXiv:0709.1260}{{\tt arXiv:0709.1260}}.

\bibitem{Schwarz:2004yj}
J.~H. Schwarz, ``{Superconformal Chern-Simons theories},'' {\em JHEP} {\bf 11}
  (2004) 078,
\href{http://www.arXiv.org/abs/hep-th/0411077}{{\tt hep-th/0411077}}.

\bibitem{VanRaamsdonk:2008ft}
M.~Van~Raamsdonk, ``{Comments on the Bagger-Lambert theory and multiple M2-
  branes},'' {\em JHEP} {\bf 05} (2008) 105,
\href{http://www.arXiv.org/abs/arXiv:0803.3803}{{\tt arXiv:0803.3803}}.

\bibitem{Aharony:2008ug}
O.~Aharony, O.~Bergman, D.~L. Jafferis, and J.~Maldacena, ``{${\cal N}=6$
  superconformal Chern-Simons-matter theories, M2-branes and their gravity
  duals},'' {\em JHEP} {\bf 10} (2008) 091,
\href{http://www.arXiv.org/abs/arXiv:0806.1218}{{\tt arXiv:0806.1218}}.

\bibitem{Aharony:2008gk}
O.~Aharony, O.~Bergman, and D.~L. Jafferis, ``{Fractional M2-branes},'' {\em
  JHEP} {\bf 11} (2008) 043,
\href{http://www.arXiv.org/abs/arXiv:0807.4924}{{\tt arXiv:0807.4924}}.

\bibitem{Klebanov:1998hh}
I.~R. Klebanov and E.~Witten, ``{Superconformal field theory on threebranes at
  a Calabi-Yau singularity},'' {\em Nucl. Phys.} {\bf B536} (1998) 199--218,
\href{http://www.arXiv.org/abs/hep-th/9807080}{{\tt hep-th/9807080}}.

\bibitem{Morrison:1998cs}
D.~R. Morrison and M.~R. Plesser, ``{Non-spherical horizons. I},'' {\em Adv.
  Theor. Math. Phys.} {\bf 3} (1999) 1--81,
\href{http://www.arXiv.org/abs/hep-th/9810201}{{\tt hep-th/9810201}}.

\bibitem{Hashimoto:1999yj}
A.~Hashimoto and N.~Itzhaki, ``{On the hierarchy between non-commutative and
  ordinary supersymmetric Yang-Mills},'' {\em JHEP} {\bf 12} (1999) 007,
\href{http://www.arXiv.org/abs/hep-th/9911057}{{\tt hep-th/9911057}}.

\bibitem{Marolf:2000cb}
D.~Marolf, ``{Chern-Simons terms and the three notions of charge},''
\href{http://www.arXiv.org/abs/hep-th/0006117}{{\tt hep-th/0006117}}.

\bibitem{Freed:1999vc}
D.~S. Freed and E.~Witten, ``{Anomalies in string theory with D-branes},''
\href{http://www.arXiv.org/abs/hep-th/9907189}{{\tt hep-th/9907189}}.

\bibitem{Bergman:2009zh}
O.~Bergman and S.~Hirano, ``{Anomalous radius shift in $AdS_4/CFT_3$},''
\href{http://www.arXiv.org/abs/arXiv:0902.1743}{{\tt arXiv:0902.1743}}.

\bibitem{Klebanov:2000hb}
I.~R. Klebanov and M.~J. Strassler, ``{Supergravity and a confining gauge
  theory: Duality cascades and $\chi$SB-resolution of naked singularities},''
  {\em JHEP} {\bf 08} (2000) 052,
\href{http://www.arXiv.org/abs/hep-th/0007191}{{\tt hep-th/0007191}}.

\bibitem{Cherkis:2002ir}
S.~A. Cherkis and A.~Hashimoto, ``{Supergravity solution of intersecting branes
  and AdS/CFT with flavor},'' {\em JHEP} {\bf 11} (2002) 036,
\href{http://www.arXiv.org/abs/hep-th/0210105}{{\tt hep-th/0210105}}.

\bibitem{Intriligator:1996ex}
K.~A. Intriligator and N.~Seiberg, ``{Mirror symmetry in three dimensional
  gauge theories},'' {\em Phys. Lett.} {\bf B387} (1996) 513--519,
\href{http://www.arXiv.org/abs/hep-th/9607207}{{\tt hep-th/9607207}}.

\bibitem{Grana:2001xn}
M.~Grana and J.~Polchinski, ``{Gauge/gravity duals with holomorphic dilaton},''
  {\em Phys. Rev.} {\bf D65} (2002) 126005,
\href{http://www.arXiv.org/abs/hep-th/0106014}{{\tt hep-th/0106014}}.

\bibitem{Polchinski:2000mx}
J.~Polchinski, ``{${\cal N} = 2$ gauge-gravity duals},'' {\em Int. J. Mod.
  Phys.} {\bf A16} (2001) 707--718,
\href{http://www.arXiv.org/abs/hep-th/0011193}{{\tt hep-th/0011193}}.

\bibitem{DiVecchia:2001uc}
P.~Di~Vecchia, H.~Enger, E.~Imeroni, and E.~Lozano-Tellechea, ``{Gauge theories
  from wrapped and fractional branes},'' {\em Nucl. Phys.} {\bf B631} (2002)
  95--127,
\href{http://www.arXiv.org/abs/hep-th/0112126}{{\tt hep-th/0112126}}.

\bibitem{Seiberg:1994pq}
N.~Seiberg, ``{Electric-magnetic duality in supersymmetric non-Abelian gauge
  theories},'' {\em Nucl. Phys.} {\bf B435} (1995) 129--146,
\href{http://www.arXiv.org/abs/hep-th/9411149}{{\tt hep-th/9411149}}.

\bibitem{Benini:2007gx}
F.~Benini, F.~Canoura, S.~Cremonesi, C.~Nunez, and A.~V. Ramallo,
  ``{Backreacting Flavors in the Klebanov-Strassler Background},'' {\em JHEP}
  {\bf 09} (2007) 109,
\href{http://www.arXiv.org/abs/arXiv:0706.1238}{{\tt arXiv:0706.1238}}.

\bibitem{Bertolini:2000dk}
M.~Bertolini {\em et al.}, ``{Fractional D-branes and their gauge duals},''
  {\em JHEP} {\bf 02} (2001) 014,
\href{http://www.arXiv.org/abs/hep-th/0011077}{{\tt hep-th/0011077}}.

\bibitem{Bertolini:2001qa}
M.~Bertolini, P.~Di~Vecchia, M.~Frau, A.~Lerda, and R.~Marotta, ``{${\cal N} =
  2$ gauge theories on systems of fractional D3/D7 branes},'' {\em Nucl. Phys.}
  {\bf B621} (2002) 157--178,
\href{http://www.arXiv.org/abs/hep-th/0107057}{{\tt hep-th/0107057}}.

\bibitem{Bertolini:2001ma}
M.~Bertolini {\em et al.}, ``{Supersymmetric 3-branes on smooth ALE manifolds
  with flux},'' {\em Nucl. Phys.} {\bf B617} (2001) 3--42,
\href{http://www.arXiv.org/abs/hep-th/0106186}{{\tt hep-th/0106186}}.

\bibitem{Aharony:2000pp}
O.~Aharony, ``{A note on the holographic interpretation of string theory
  backgrounds with varying flux},'' {\em JHEP} {\bf 03} (2001) 012,
\href{http://www.arXiv.org/abs/hep-th/0101013}{{\tt hep-th/0101013}}.

\bibitem{Witten:1997sc}
E.~Witten, ``{Solutions of four-dimensional field theories via M-theory},''
  {\em Nucl. Phys.} {\bf B500} (1997) 3--42,
\href{http://www.arXiv.org/abs/hep-th/9703166}{{\tt hep-th/9703166}}.

\bibitem{Seiberg:1996nz}
N.~Seiberg and E.~Witten, ``{Gauge dynamics and compactification to three
  dimensions},''
\href{http://www.arXiv.org/abs/hep-th/9607163}{{\tt hep-th/9607163}}.

\bibitem{Johnson:1999qt}
C.~V. Johnson, A.~W. Peet, and J.~Polchinski, ``{Gauge theory and the excision
  of repulson singularities},'' {\em Phys. Rev.} {\bf D61} (2000) 086001,
\href{http://www.arXiv.org/abs/hep-th/9911161}{{\tt hep-th/9911161}}.

\bibitem{Benini:2008ir}
F.~Benini, M.~Bertolini, C.~Closset, and S.~Cremonesi, ``{The ${\cal N}=2$
  cascade revisited and the enhancon bearings},''
\href{http://www.arXiv.org/abs/arXiv:0811.2207}{{\tt arXiv:0811.2207}}.

\bibitem{ori_branes}
O.~Ganor.
\newblock unpublished.

\bibitem{Hanany:1996ie}
A.~Hanany and E.~Witten, ``{Type IIB superstrings, BPS monopoles, and three-
  dimensional gauge dynamics},'' {\em Nucl. Phys.} {\bf B492} (1997) 152--190,
\href{http://www.arXiv.org/abs/hep-th/9611230}{{\tt hep-th/9611230}}.

\bibitem{Gaiotto:2009tk}
D.~Gaiotto and D.~L. Jafferis, ``{Notes on adding D6 branes wrapping $RP^3$ in
  $AdS_4 \times CP^3$},''
\href{http://www.arXiv.org/abs/arXiv:0903.2175}{{\tt arXiv:0903.2175}}.

\bibitem{Hikida:2009tp}
Y.~Hikida, W.~Li, and T.~Takayanagi, ``{ABJM with Flavors and FQHE},''
\href{http://www.arXiv.org/abs/arXiv:0903.2194}{{\tt arXiv:0903.2194}}.

\bibitem{Fujita:2009xz}
M.~Fujita and T.-S. Tai, ``{Eschenburg space as gravity dual of flavored ${\cal
  N}$=4 Chern-Simons-matter theory},''
\href{http://www.arXiv.org/abs/arXiv:0906.0253}{{\tt arXiv:0906.0253}}.

\bibitem{Witten:2009xu}
E.~Witten, ``{Branes, Instantons, And Taub-NUT Spaces},''
\href{http://www.arXiv.org/abs/arXiv:0902.0948}{{\tt arXiv:0902.0948}}.

\bibitem{Kitao:1998mf}
T.~Kitao, K.~Ohta, and N.~Ohta, ``{Three-dimensional gauge dynamics from brane
  configurations with $(p,q)$-fivebrane},'' {\em Nucl. Phys.} {\bf B539} (1999)
  79--106,
\href{http://www.arXiv.org/abs/hep-th/9808111}{{\tt hep-th/9808111}}.

\bibitem{Bergman:1999na}
O.~Bergman, A.~Hanany, A.~Karch, and B.~Kol, ``{Branes and supersymmetry
  breaking in 3D gauge theories},'' {\em JHEP} {\bf 10} (1999) 036,
\href{http://www.arXiv.org/abs/hep-th/9908075}{{\tt hep-th/9908075}}.

\bibitem{Dasgupta:1999wx}
K.~Dasgupta and S.~Mukhi, ``{Brane constructions, fractional branes and anti-de
  Sitter domain walls},'' {\em JHEP} {\bf 07} (1999) 008,
\href{http://www.arXiv.org/abs/hep-th/9904131}{{\tt hep-th/9904131}}.

\bibitem{Bachas:1997sc}
C.~Bachas and M.~B. Green, ``{A classical manifestation of the Pauli exclusion
  principle},'' {\em JHEP} {\bf 01} (1998) 015,
\href{http://www.arXiv.org/abs/hep-th/9712187}{{\tt hep-th/9712187}}.

\bibitem{Bachas:1997kn}
C.~P. Bachas, M.~B. Green, and A.~Schwimmer, ``{(8,0) quantum mechanics and
  symmetry enhancement in type I' superstrings},'' {\em JHEP} {\bf 01} (1998)
  006,
\href{http://www.arXiv.org/abs/hep-th/9712086}{{\tt hep-th/9712086}}.

\bibitem{Lee:1996kz}
K.-M. Lee, E.~J. Weinberg, and P.~Yi, ``{The moduli space of many BPS monopoles
  for arbitrary gauge groups},'' {\em Phys. Rev.} {\bf D54} (1996) 1633--1643,
\href{http://www.arXiv.org/abs/hep-th/9602167}{{\tt hep-th/9602167}}.

\bibitem{Gauntlett:1997pk}
J.~P. Gauntlett, G.~W. Gibbons, G.~Papadopoulos, and P.~K. Townsend,
  ``{Hyper-K\"ahler manifolds and multiply intersecting branes},'' {\em Nucl.
  Phys.} {\bf B500} (1997) 133--162,
\href{http://www.arXiv.org/abs/hep-th/9702202}{{\tt hep-th/9702202}}.

\bibitem{Hashimoto:2008iv}
A.~Hashimoto and P.~Ouyang, ``{Supergravity dual of Chern-Simons Yang-Mills
  theory with ${\cal N}=6,8$ superconformal IR fixed point},'' {\em JHEP} {\bf
  10} (2008) 057,
\href{http://www.arXiv.org/abs/arXiv:0807.1500}{{\tt arXiv:0807.1500}}.

\bibitem{Witten:1996md}
E.~Witten, ``{On flux quantization in M-theory and the effective action},''
  {\em J. Geom. Phys.} {\bf 22} (1997) 1--13,
\href{http://www.arXiv.org/abs/hep-th/9609122}{{\tt hep-th/9609122}}.

\bibitem{Sethi:1998zk}
S.~Sethi, ``{A relation between ${\cal N} = 8$ gauge theories in three
  dimensions},'' {\em JHEP} {\bf 11} (1998) 003,
\href{http://www.arXiv.org/abs/hep-th/9809162}{{\tt hep-th/9809162}}.

\bibitem{Green:1996dd}
M.~B. Green, J.~A. Harvey, and G.~W. Moore, ``{I-brane inflow and anomalous
  couplings on D-branes},'' {\em Class. Quant. Grav.} {\bf 14} (1997) 47--52,
\href{http://www.arXiv.org/abs/hep-th/9605033}{{\tt hep-th/9605033}}.

\bibitem{Eguchi:1980jx}
T.~Eguchi, P.~B. Gilkey, and A.~J. Hanson, ``{Gravitation, Gauge Theories and
  Differential Geometry},'' {\em Phys. Rept.} {\bf 66} (1980)
213.

\bibitem{Witten:1998xy}
E.~Witten, ``{Baryons and branes in anti de Sitter space},'' {\em JHEP} {\bf
  07} (1998) 006,
\href{http://www.arXiv.org/abs/hep-th/9805112}{{\tt hep-th/9805112}}.

\bibitem{Ohta:1999iv}
K.~Ohta, ``{Supersymmetric index and $s$-rule for type IIB branes},'' {\em
  JHEP} {\bf 10} (1999) 006,
\href{http://www.arXiv.org/abs/hep-th/9908120}{{\tt hep-th/9908120}}.

\bibitem{Strassler:2005qs}
M.~J. Strassler, ``{The duality cascade},''
\href{http://www.arXiv.org/abs/hep-th/0505153}{{\tt hep-th/0505153}}.

\bibitem{Dymarsky:2005xt}
A.~Dymarsky, I.~R. Klebanov, and N.~Seiberg, ``{On the moduli space of the
  cascading $SU(M+p) \times SU(p)$ gauge theory},'' {\em JHEP} {\bf 01} (2006)
  155,
\href{http://www.arXiv.org/abs/hep-th/0511254}{{\tt hep-th/0511254}}.

\bibitem{nextgreatpaper}
O.~Aharony, A.~Hashimoto, S.~Hirano, and P.~Ouyang.
\newblock work in progress.

\bibitem{Russo:1996if}
J.~G. Russo and A.~A. Tseytlin, ``{Waves, boosted branes and BPS states in
  M-theory},'' {\em Nucl. Phys.} {\bf B490} (1997) 121--144,
\href{http://www.arXiv.org/abs/hep-th/9611047}{{\tt hep-th/9611047}}.

\bibitem{Kachru:2002sk}
S.~Kachru, M.~B. Schulz, P.~K. Tripathy, and S.~P. Trivedi, ``{New
  supersymmetric string compactifications},'' {\em JHEP} {\bf 03} (2003) 061,
\href{http://www.arXiv.org/abs/hep-th/0211182}{{\tt hep-th/0211182}}.

\bibitem{Shelton:2005cf}
J.~Shelton, W.~Taylor, and B.~Wecht, ``{Nongeometric Flux Compactifications},''
  {\em JHEP} {\bf 10} (2005) 085,
\href{http://www.arXiv.org/abs/hep-th/0508133}{{\tt hep-th/0508133}}.

\end{thebibliography}\endgroup

\end{document}